\documentclass[10pt]{iopart}

\usepackage[square,numbers,sort&compress]{natbib}
\expandafter\let\csname equation*\endcsname\relax
\expandafter\let\csname endequation*\endcsname\relax

\usepackage{graphicx}
\graphicspath{ {figures/} }
\usepackage{DC_header}

\begin{document}

\title[Properties of correlated fission fragments from neutron induced fission of Np-237]{Properties of correlated fission fragments from neutron induced fission of Np-237 at incident neutron energies between 200 keV and 100 MeV}

\author{Devin Connolly$^1$,
K.~B.~Montoya$^{1,2}$,
D.~L.~Duke$^1$,
U.~Greife$^2$,
A.~E.~Lovell$^1$,
S.~Mosby$^1$,
C.~Prokop$^1$,
E.~Rudziensky$^1$,
K.~Schmitt$^1$\footnote{Present Address: Oak Ridge National Laboratory, Oak Ridge, TN},
J.~Winkelbauer$^1$}

\address{$^1$ Los Alamos National Laboratory, Los Alamos, New Mexico 87545, USA}
\address{$^2$ Colorado School of Mines, Golden, CO, 80401, USA}
\ead{dconnolly@lanl.gov}

\begin{abstract}
Neutron-induced fission of \np{237}\ has been measured over a wide range of incident neutron energies using a twin Frisch-gridded ionization chamber (TFGIC) and a thin-backed \np{237} target. These measurements were performed at the Los Alamos Neutron Science Center - Weapons Neutron Research (LANSCE - WNR) facility, which provides a collimated beam of neutrons with energies ranging from 100s of keV to 100s of MeV. The data were analyzed using the double-energy ($2E$) method, with mass-dependent corrections for prompt-fission neutrons and pulse height defect. Pre- and post-neutron evaporation average total kinetic energy (\tke) values are reported for 54 incident neutron energies in the energy range $0.20 \le E_n \le 100.0$~MeV and compared to existing data and evaluations. Pre- and post-neutron evaporation mass yields were extracted with a full width at half maximum (FWHM) resolution of 4u and compared to existing data and evaluations. The present \tke\ and mass yield data agree with previous results and also with statistical models of \npnf\ at incident neutron energies between $E_n = 0.2 - 20.0$~MeV. A flattening of the \tke\ data is observed (relative to the prediction of the GEF model) above $E_n = 20.0$~MeV. However, the interpretation of this discrepancy is unclear as the analysis method's neglect of incomplete momentum transfer at high energies, as well as pre-equilibrium pre-fission phenomena likely have a significant impact on the measurement at such high incident neutron energies.
\end{abstract}

\submitto{\jpg}
\noindent{\it Keywords\/}: actinides, fission, fission mass yields, neutron-induced fission, total kinetic energy release
\maketitle

\section{Introduction\label{sec:intro}}
Measurements of the neutron-induced fission of major actinides are vital to applications such as nuclear energy, nuclear forensics, nuclear nonproliferation and stockpile stewardship. Contemporary fission models are informed by additional correlated fission data, advancing toward a fully predictive fission model. Of particular interest are the average total kinetic energy (\tke) of the fission fragments, fission fragment mass yield distributions and correlations between mass and \tke\ as they evolve with excitation energy of the fissioning system. Neptunium-237 is produced in significant quantities in the nuclear fuel cycle. Its presence in spent reactor fuel and $\sim2\times10^6$ year half life make a complete and precise understanding of its production and destruction desirable to inform reactor design. Little \tke\ data exists for \npnf, and until recently, there was no \tke\ nor mass yield data for \npnf\ at incident neutron energies above $E_n = 5.55\,\mathrm{MeV}$. The most recent measurement suffered from low statistics and relatively poor energy resolution compared to previous studies, so a higher quality data set is desired. 
The majority of the energy released in fission is contained the kinetic energy of the fission fragments as they accelerate away from each other following scission. The average total kinetic energy \tke\ is the sum of the kinetic energies of the two fragments averaged over all fragment mass splits. This quantity varies with the excitation energy of the nucleus which in turn depends on the incident neutron energy. Further, the distribution of fission fragment masses also evolves with the excitation energy of the nucleus. Many years of fission research have resulted in a library of nuclear data useful for fission applications \cite{, Brown2018,England1993}. New advances in fission modeling and applications have revealed a need for additional data at higher incident neutron energies than is currently available and in general, post-scission fission observables data from neutron induced fission of actinides is sparse \cite{Madland2006}. In particular, data on neutron-induced fission of \np{237} are lacking at high incident neutron energies.\\ 
\indent Neutron induced fission of \np{237} provides a probe of the nuclear structure of fissioning systems. Because the compound nucleus (\np{238}) is an odd-odd fissioning system, the study of \npnf\ complements previous studies of even-odd (e.g.~\u{235,9}) and even-even (e.g.~\pu{240}, \u{234,6}) fissioning systems. Furthermore, \npnf\ is of interest to energy and national security applications, as \np{237} is produced in significant quantities in the nuclear fuel cycle. Its presence in spent reactor fuel along with its $2.14\times10^6$ year half life and its nature as a fissionable material make a complete and precise understanding of its production and destruction desirable to inform reactor design, particularly that of fast reactors. Study of \npnf\ also provides vital information for stockpile stewardship, nuclear forensics and nonproliferation applications. \\
\indent This paper presents results on pre- and post-neutron evaporation \tke\ and mass yields from correlated fission products of \npnf\ measured using a twin Frisch-gridded ionization chamber and the unmoderated white neutron source at LANSCE-WNR. \Cref{sec:prevMeas} discusses previous studies on \npnf\ in detail. Experimental methods, devices and facilities are detailed in \cref{sec:Experiment}. Data analysis methods are presented in \cref{sec:Analysis} and results are presented in \cref{sec:Results}.
\begin{figure}[t]
\begin{indented}
  \item[] \includegraphics[trim={0.25cm 0.125cm 0.375cm 0.25cm}, clip,width=0.6\textwidth]{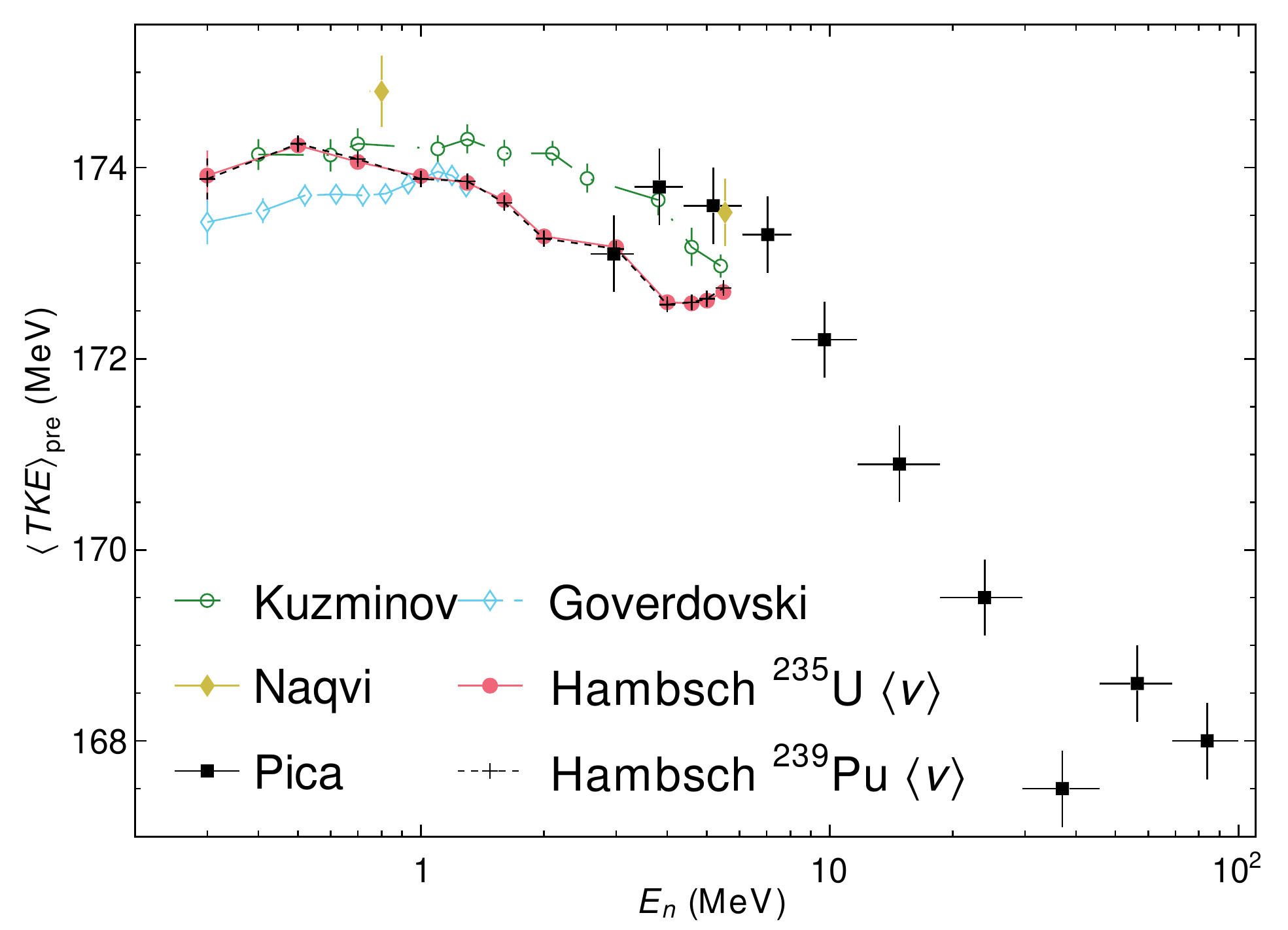}
\end{indented}
  \caption{\label{fig:np237lit} Existing pre-neutron evaporation \npnf\ \tke\ vs $E_{n}$ data
\cite{Kuzminov1970, Naqvi1986, Goverdovski1992, Hambsch2000, Pica2020} in the range $E_n = 0 - 6.5$~MeV. Data from references \cite{Bennett1967, Asghar1977, Ruiz1978, Wagemans1981} excluded from this plot. Data from Kuzminov \cite{Kuzminov1970} and Goverdovski \cite{Goverdovski1992} have been renormalized using the currently accepted value of $\langle TKE\rangle_{\mathrm{pre}}$ for thermal neutron induced fission of ${}^{235}$U. See text for details.}
\end{figure}
\section{\label{sec:prevMeas}Previous measurements}
Although the fission cross section of \np{237} has been measured over a wide range of incident neutron energies, little \tke\ data exists for \npnf\ \cite{Bennett1967, Kuzminov1970, Asghar1977, Ruiz1978, Wagemans1981, Naqvi1986, Goverdovski1992, Hambsch2000} and until a recent study by Pica \etal\ \cite{Pica2020}, there was no \tke\ nor mass yield data for incident neutron energies above $E_n = 5.55\,\mathrm{MeV}$ (see \cref{fig:np237lit}). Due to the variety of the details of the previous studies, some specificity in unpacking and synthesizing the available information is necessary. For clarity, the details of each study will be discussed individually in the following sections.\\
\indent Previous studies of \npnf\ have utilized one of two experimental techniques to measure \tke\ and fission mass yields. The most commonly employed technique was that of an evacuated scattering chamber containing a pair of solid-state silicon detectors (of varying type) mounted collinearly on opposite sides of a thin-backed \np{237} target with various neutron sources incident on the target \cite{Bennett1967, Kuzminov1970, Asghar1977, Ruiz1978, Wagemans1981, Muller1981, Naqvi1986, Goverdovski1992, Pica2020}. The target was mounted at an oblique angle with respect to the incident neutron beam and the detectors were mounted at an oblique angle with respect to both the target and neutron beam. The Si detectors measured the energy of coincident fission products, and (with the exception of reference \cite{Naqvi1986} - see \cref{sec:Naqvi}) the $2E$ method was used to analyze the data. The second technique was the same technique used in the present work - a twin Frisch-gridded ionization chamber (TFGIC) containing a thin-backed \np{237} target was used to measure the energies and polar angles of coincident fission products and the $2E$ method was used to analyze the data.
\subsection{Bennett and Stein}
Bennett and Stein \cite{Bennett1967} used the first of the two experimental techniques described above. Two silicon surface barrier (SSB) detectors were used to detect coincident fission products. The SSB detectors were collimated by annular diaphragms with rounded edges in order to avoid edge effects. The \np{237} target was irradiated in a reactor beamline by fission spectrum neutrons (viz.~neutrons ranging in energy from $\sim0.01 - 10$~MeV - cf.\ reference \cite{Neudecker2018}), with an interceding boron absorber to suppress thermal neutrons. The pulse height spectra of the SSBs were energy calibrated and corrected for pulse height defect (PHD) using the Schmitt method \cite{Schmitt1965} and a regular reference measurement of thermal neutron induced fission of \u{235}. Although the \tke\ datum of $174.0\pm2.0$~MeV obtained by Bennett and Stein is consistent with all subsequent measurements, the absence of an unambiguous correlation to neutron energy and relatively broad uncertainty on this datum render it to be of limited utility in interpreting the present data. It has therefore been excluded from \cref{fig:np237lit} and will not be discussed further.
\subsection{Kuz'minov \etal}
Kuz'minov \etal\ similarly used a pair of SSB detectors situated collinearly on opposite sides of a thin \np{237} target to detect coincident fission products \cite{Kuzminov1970}. The pulse height spectra of the SSBs were energy calibrated and corrected for PHD using the Schmitt method with a reference measurement of \unthf\ using a value of $\tkeunth = 171.9\pm1.4$~MeV \cite{Schmitt1966}. No mention is made in reference \cite{Kuzminov1970} nor the references therein of the use of diaphragms to avoid edge effects. Monoenergetic neutrons were produced by \pnrxn{}{T}{3}{He} and \dnrxn{}{D}{3}{He} reactions using a Van de Graaff accelerator. Mass yield and \tke\ data were collected for nine different incident neutron energies ranging from $E_n = 0.7 - 5.4$~MeV. The data were analyzed using the $2E$ method. Because this data is not available in EXFOR \cite{Otuka2014, exfor}, the data presented in \cref{fig:np237lit} was digitized from the original source and renormalized using the most current recommended value of $\tkeunth = 170.5\pm0.5$~MeV \cite{Wagemans1991, Bertsch2015}.
\subsection{Asghar \etal}
Asghar \etal\ used a pair of SSB detectors situated collinearly on opposite sides of a thin \np{237} target to detect coincident fission products from \npnthf\ \cite{Asghar1977}. The SSBs were collimated with diaphragms in order to avoid edge effects. The pulse height spectra of the SSBs were energy calibrated and corrected for PHD using the Schmitt method with a reference measurement of \unthf. The neutron source was a curved, collimated beamline of a high flux reactor with a high ratio of thermal to fast neutrons. The data were analyzed using the $2E$ method. Asghar \etal\ measured a post-neutron evaporation \tke\ value of $\left<TKE\right>_{n_{\mathrm{th}}}^{(\mathrm{post})} = 170.7\pm0.7$~MeV and calculated a pre-neutron evaporation \tke\ value of $\left<TKE\right>_{n_{\mathrm{th}}}^{(\mathrm{pre})} = 172.4\pm0.7$~MeV. These values are inconsistent with previous and subsequent measurements. In particular, Thierens \etal\ \cite{Thierens1980} obtained results suggesting that the Asghar results were erroneous and Wagemens \etal\ \cite{Wagemans1981} contradicted the Asghar \etal\ result using the same experimental technique at the same facility (see section \cref{sec:Wagemans}). Goverdovski and Mitrofanov \cite{Goverdovski1992} later cast further doubt on this measurement, pointing out that the target used by Asghar \etal\ likely had significant ${}^{239}$Pu contamination. For these reasons, this datum has been excluded from \cref{fig:np237lit} and will not be discussed further.
\subsection{Ruiz}
Ruiz used a pair of SSB detectors situated collinearly on opposite sides of a thin \np{237} target to detect coincident fission products \cite{Ruiz1978}. Quasi-monoenergetic neutrons were produced by \pnrxn{7}{Li}{7}{Be} using a Van de Graaff accelerator. The details in EXFOR state that ``the neutron energy spread is 30 keV at 500 keV,'' presumably referring to the full width at half maximum (FWHM) of the neutron energy distribution. The SSBs pulse height spectra were calibrated and corrected for PHD using the Schmitt method with a reference measurement of \unthf. The data were analyzed using the 2E method and \tke\ values were reported for six different incident neutron energies ranging from $E_n = 0.5 - 1.75$~MeV, with a measurement at $E_n = 0.75$~MeV being repeated twice. Although this data is published in EXFOR, the original source is a Ph.D.\ thesis (viz.\ not peer-reviewed) from l'Universit\'e de Bordeaux in 1978 and could not be obtained by the authors. Although various details of the experiment are published in EXFOR \cite{exfor}, no uncertainty quantification nor error analysis details are given (with the exception of the quoted spread in $E_n$), and the data are reported without error bars. The experiment evidently suffered from a significant systematic offset as the data are lower than the literature by $\approx7.5$~MeV on average. Furthermore, there are large fluctuations between adjacent data points and since there are no error bars, it is impossible to discern if the fluctuations can be attributed to statistical fluctuations or if they are indicative of poor data quality. For these reasons, the authors have deemed this data to be of limited utility in interpreting the present data and as such it has been excluded from \cref{fig:np237lit} and will not be discussed further.
\subsection{Wagemans \etal}\label{sec:Wagemans}
Wagemans \etal\ used a pair of SSB detectors situated collinearly on opposite sides of a thin \np{237} target to detect coincident fission products from \npnthf\ \cite{Wagemans1981}. The SSBs were collimated with diaphragms in order to avoid edge effects. The pulse height spectra of the SSBs were energy calibrated and corrected for PHD using the Schmitt method with a reference measurement of \unthf. The neutron source was a curved, collimated beamline of a high flux reactor with a high ratio of thermal to fast neutrons. The data were analyzed using the $2E$ method. Wagemans \etal\ measured a post-neutron evaporation \tke\ value of $\tke_{\mathrm{post}} = 174.7\pm0.6$~MeV and calculated a pre-neutron evaporation \tke\ value of $\tke_{\mathrm{pre}} = 176.4\pm0.6$~MeV. However, Wagemens \etal\ used a value of $\tkeunth = 172.7\pm0.5$~MeV as their energy calibration reference. Therefore, the datum in \cref{fig:np237le} has been renormalized to the current recommended value of $\tkeunth = 170.5\pm0.5$~MeV, yielding respective pre- and post-neutron emission \tke\ values of $\tke_{\mathrm{pre}} = 174.2\pm0.6$~MeV and $\tke_{\mathrm{post}} = 172.5\pm0.6$~MeV for thermal neutron induced fission of \np{237}.
\subsection{Naqvi \etal and M\"{u}ller \etal\label{sec:Naqvi}}
The experimental technique employed by Naqvi \etal\ \cite{Naqvi1986} (described in detail in reference \cite{Muller1984}) was a ``double-energy, double-velocity'' or $2E-2v$ measurement. To date, this is the only $2E-2v$ measurement of \npnf. The $2E-2v$ method enables calculation of the post-neutron emission masses directly from the measured energies and velocities of the products using their (non-relativistic) kinetic energy:
\begin{equation}\label{eqn:KE}
    m_{0,1} = \frac{2E_{0,1}}{v_{0,1}^2}
\end{equation}
Additionally, if one makes the approximation that the prompt fission neutrons are emitted isotropically and originate from fully accelerated fragments, then the velocities of the fission products are equal to the velocities of the primary fragments (on average) and one may extract the neutron multiplicity $\langle\nu\rangle$ as a function of the primary fragment mass via
\begin{equation}
    \nu\left(m_{0,1}^*\right) = \frac{\displaystyle M_{\mathrm{CN}}v_{1,0}  - m_{0,1}^*\left(v^*_0 + v^*_1\right)} {\displaystyle m_n\left(v^*_0 + v^*_1\right)}
\end{equation}
\indent The experimental technique employed was the first technique described in \cref{sec:prevMeas}. A thin-backed \np{237} target was mounted at an oblique angle with respect to an incident neutron beam. Quasi-monoenergetic neutrons were produced by \pnrxn{7}{Li}{7}{Be} and \dnrxn{}{D}{3}{He} using a pulsed Van de Graaff accelerator. The accelerator was pulsed at repetition rate of 5MHz with a 700~ps pulse width and a beam pickup detector was used as the time-zero reference for the ToF measurement. A pair of collinear, large-area SSB detectors mounted orthogonally to the \np{237} target were used to measure the energies and times of flight (with respect to the $T_0$) of coincident fission products. Various diaphragms in the vacuum chamber prevented detection of scattered fission products. The pulse height spectra of the SSBs was calibrated with regular measurements of \cf{252} spontaneous fission spectra. The \cf{252} sources were installed on a retractable drive inside the vacuum chamber and were produced via self-transfer. Consequently, the calibration sources had negligible thickness. Pre-neutron evaporation \tke\ and mass yields were reported for two different incident neutron energies. Additionally, Naqvi \etal\ reported the average neutron multiplicity as a function of primary fragment mass \nubar\ for both incident neutron energies (note that although the \tke\ values quoted in this manuscript are from reference \cite{Naqvi1986}, the \nubar\ data was retrieved from EXFOR \cite{exfor} and are from reference \cite{Muller1981}). Because of the measurement technique and calibration method employed by Naqvi \etal, Hambsch \etal\ \cite{Hambsch2000} note that the data of Naqvi \etal\ give ``...absolute \tke\ values via the measurement of the fragment velocity.''
\subsection{Goverdovskii and Mitrofanov\label{sec:Goverdovskii}}
Goverdovskii and Mitrofanov also used a pair of SSB detectors situated collinearly on opposite sides of a thin \np{237} target to detect coincident fission products \cite{Goverdovski1992}. The pulse height spectra of the SSBs were energy calibrated and corrected for PHD using the Schmitt method with a reference measurement of \unthf. No mention is made in reference \cite{Goverdovski1992} nor the references therein of the use of diaphragms to avoid edge effects. Quasi-monoenergetic neutrons were produced by \pnrxn{}{T}{3}{He} reaction using a Cockcroft-Walton generator. Mass yield and \tke\ data were reported for ten different incident neutron energies ranging from $E_n = 0.28 - 1.28$~MeV. The data were analyzed using the $2E$ method. Because this data is not available in EXFOR, the data presented in \cref{fig:np237lit} was digitized from the original source and renormalized using using the most current recommended value of $\tkeunth = 170.5\pm0.5$~MeV.
\subsection{Hambsch \etal\label{sec:Hambsch}}
Hambsch \etal is the only study of \npnf\ (aside from the present study) to use a TFGIC to measure \tke\ and FPY. Monoenergetic neutrons were produced by the \pnrxn{7}{Li}{7}{Be}, \pnrxn{}{T}{3}{He} and \dnrxn{}{D}{3}{He} reactions using a Van de Graaff accelerator. The pulse height spectra of the TFGIC anode and grid signals were energy calibrated with a reference measurement of \unthf\ using a value of $\tkeunth = 170.5\pm0.5$~MeV \cite{Wagemans1991, Bertsch2015}. The data were analyzed using the $2E$ method and Mass yield and \tke\ data were reported for twelve different incident neutron energies in the range $E_n = 0.3 - 5.5$ MeV. Additionally, Hambsch \etal\ used both $^{235}\mathrm{U}$ and \pu{239}\ neutron emission distributions ($\langle\nu\rangle(m^*)$ - see \cref{sec:nubar}) in the $2E$ analysis, demonstrating that $\langle TKE\rangle_{\mathrm{pre}}$ only weekly depends on $\langle \nu\rangle (m^*)$. The \tke\ results obtained with both \u{235}\ and \pu{239} $\langle \nu\rangle (m^*)$ are shown in \cref{fig:np237lit,fig:np237le} (red full circles and black crosses, respectively).
\subsection{Pica \etal}
Pica \etal used a pair of $2\times2$ Si PIN diode detector arrays mounted collinearly on opposite sides of a thin backed \np{237} target \cite{Pica2020} to detect coincident fission products. The measurement was performed at the same facility as the present work (described in \ref{sec:wnr}), but at a different flight path. The characteristics of that flight path are similar enough to that described in \ref{sec:wnr} as to warrant no further discussion here. Pica \etal\ make no mention of any type of collimation of the detector arrays nor the flight tubes of the chamber to prevent the detection of scattered fission fragments or edge effects, nor do references \cite{King2017, King2018, Yanez2018}. Pica \etal\ make no mention of the method of energy calibration that was used for the Si PIN detector arrays, noting that ``The pulse height defect of the detectors for fission fragments was determined by applying the Schmitt method to the known fission spectra of \cf{252}.'' Pica \etal\ report pre- and post-neutron evaporation \tke\ values and mass yields for ten different incident neutron energies ranging from (mean) neutron energy values of $E_n = 2.96 - 83.8$~MeV. \\
\indent The incident neutron energy binning for these data is quite broad ($\sim0.7$~MeV at $E_n = 2.96$~MeV and increasing to $>31$~MeV at $E_n = 83.8$~MeV) and the uncertainty on the \tke\ values is relatively large compared to previous studies. This is likely the result of low statistics as well as the relatively poor energy resolution of the detectors used. The (low energy) pre-neutron evaporation \tke\ values reported by Pica \etal\ are compared with previous literature data in \cref{fig:np237lit}.
\section{Experiment \label{sec:Experiment}}
Particulars of the experiment including the beam facility, detector, data acquisition system and target are detailed in the following sections.
\subsection{Neutron Source \label{sec:wnr}}
The measurements were performed at the Los Alamos Neutron Science Center (LANSCE) Weapons Neutron Research (WNR) facility. A pulsed beam of 800~MeV protons incident on an unmoderated tungsten spallation target produces neutrons feeding multiple flight paths with energies ranging from 100s of keV to 100s of MeV. The LANSCE WNR facility is described in detail in references \cite{Lisowski1990, Lisowski2006}. The pulse structure of the proton beam consists of a 625~$\mu$s macropulse with a 60~Hz repetition rate. Each macropulse is composed of 125~ps (FWHM) micropulses separated by 1.8~$\mu$s. Using timing information from the accelerator and detector, the neutron energies can be determined using the neutron time-of-flight (TOF) method. 
\subsection{Detector and DAQ\label{sec:detector}}
\begin{figure}[t]
\begin{indented}
  \item[] \includegraphics[clip, width=0.5\textwidth]{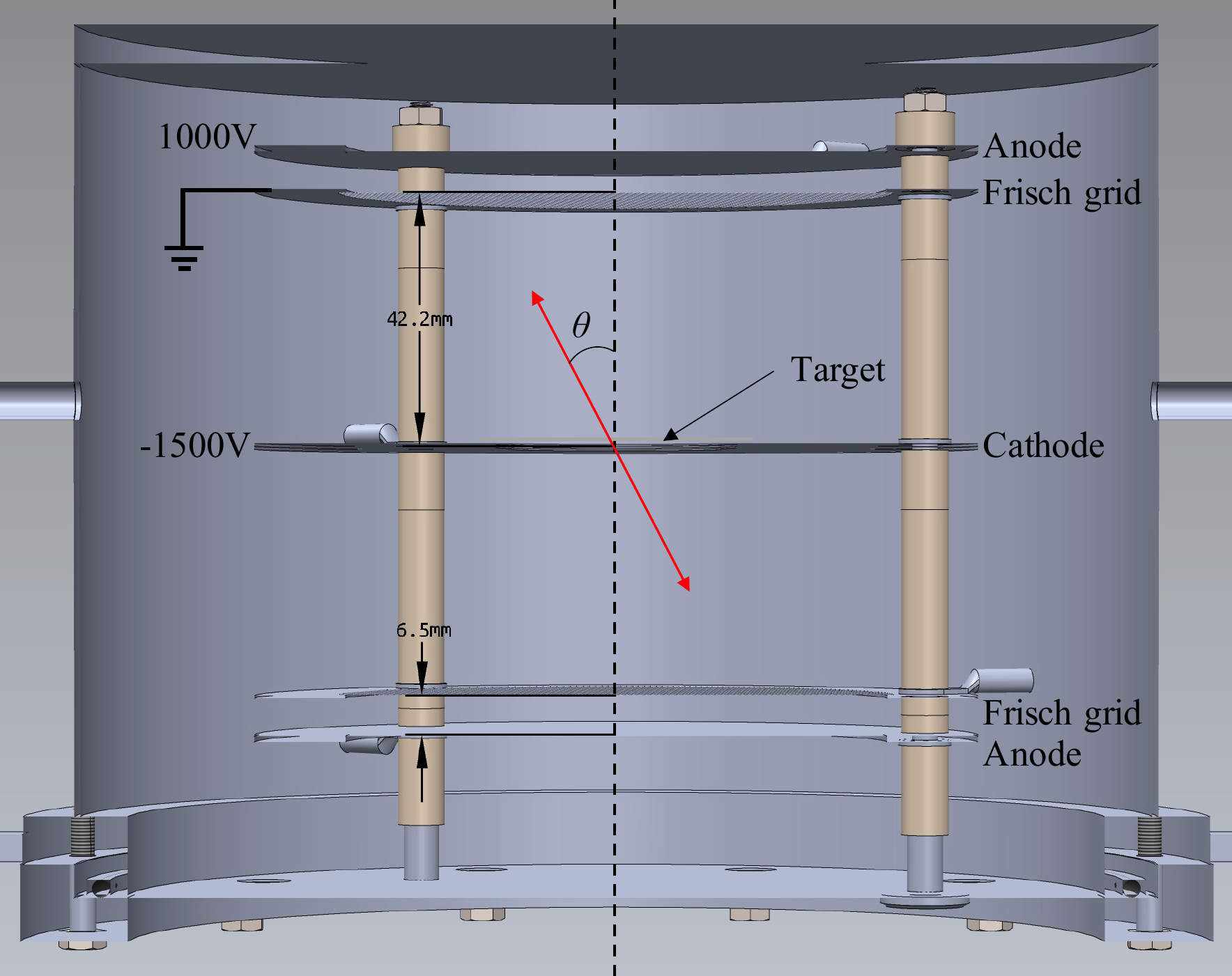}
\end{indented}
  \caption{\label{fig:tfgic} Schematic of the twin Frisch-gridded ionization chamber (TFGIC) detector used to detect coincident fission fragments from \npnf.}
\end{figure}

A twin Frisch-gridded ionization chamber (TFGIC) was used to perform the present measurements. The use of a TFGIC to measure correlated fission products has many advantages over that of Si charged particle detectors. It has a high efficiency, covering a solid angle of nearly 4$\pi$~sr as well as superior energy resolution, typically in the range of $\sim0.5 - 1$~MeV for fission fragments \cite{Budtz-Joergensen1987}. TFGIC detectors have been used in many double-energy style measurements in the past \cite{Straede1987, Vives2000, Hambsch2000, Birgersson2007, Duke2016, Meierbachtol2016, Higgins2020}.\\
\indent The TFGIC detector consists of two anodes, two wire grids, and a central aluminum cathode that are biased to electrical potentials of 1000V, ground (not biased), and -1500V respectively and is identical to the detectors described in \cite{Vives1998, Duke2016}. The anode-grid distance is $6.5\pm0.9$~mm, the cathode-grid distance is $42.2\pm 0.9$~mm and the grid wires have a $0.1\pm0.01$~mm diameter with $1\pm0.01$~mm pitch. A schematic of the detector is shown in \cref{fig:tfgic}. The detector was filled with P-10 gas (composed of 90\% argon and 10\% methane), which was continuously flowed through the chamber at a rate of $\leq50$~sccm and held at a constant pressure of $950\pm1$~mbar via a mass flow controller.\\
\indent A thin \np{237} target was mounted in the center of the cathode plane, directly in the center of the detector. Fission products are emitted back-to-back in the detector and ionize the gas as they traverse the volume between the cathode and grid. The ionized electrons induce a signal on the anode after drifting across the chamber and through the Frisch-grids, which shield the anodes from the charge induced by the electrons until the electrons pass them. The energy deposited on the anodes is proportional to the energy of the fission fragments and the ratio of the grid to anode signals is proportional to the polar angle of their trajectory.\\
\indent The charge induced on the two grids and two anodes produces analog signals that were read out using MesyTec MPR-1 charge sensitive preamplifiers (CSP). The cathode signal which is much faster (with a rise time on the order of nanoseconds), was read out by a custom fast timing preamplifier. The preamplifier signals were digitized using a CAEN VX1720 12 bit, 250 MS/s waveform digitizer. The data acquisition system is described in detail in reference \cite{Mosby2014} and waveform analysis methods are discussed in \cref{sec:WaveformProcessing}.
\subsection{Target}
A thin-backed \np{237} target was used. The target consisted of a $56.0\pm5.6~\mu \mathrm{g}/\mathrm{cm}^2$ layer of NpO$_2$ deposited onto a $100\pm10~\mu \mathrm{g}/\mathrm{cm}^2$ C backing using a molecular plating technique \cite{Loveland2009}. The \np{237} target area had a diameter of $\sim1$~cm.
\section{Analysis \label{sec:Analysis}}
The following sections describe in detail the methods and techniques utilized to extract physics information from the raw data. Section \ref{sec:WaveformProcessing} details the methods used to extract pulse height spectra and timing information from the raw waveform data. Section \ref{sec:nTof} describes the determination of the incident neutron energy and calibration method. Section \ref{sec:aphCorr} details corrections made to the anode pulse height data for effects arising from the experimental technique employed, including corrections for angular distribution, grid inefficiency, neutron momentum, energy loss and gain corrections. Section \ref{sec:eCal} describes energy calibration of the anode pulse height spectra. Finally, section \ref{sec:2E} details the $2E$ method (see section \cref{sec:nTof}).
\subsection{Waveform Processing \label{sec:WaveformProcessing}}
\begin{figure}[t]
\begin{indented}
\item[]
\begin{tikzpicture}[scale=1]
  \begin{scope}
    \node[anchor=south west, inner sep=0](wf) at (0,0) {\includegraphics[trim={0.25cm 0.125cm 0.25cm 0.125cm}, clip, width=0.67\textwidth]{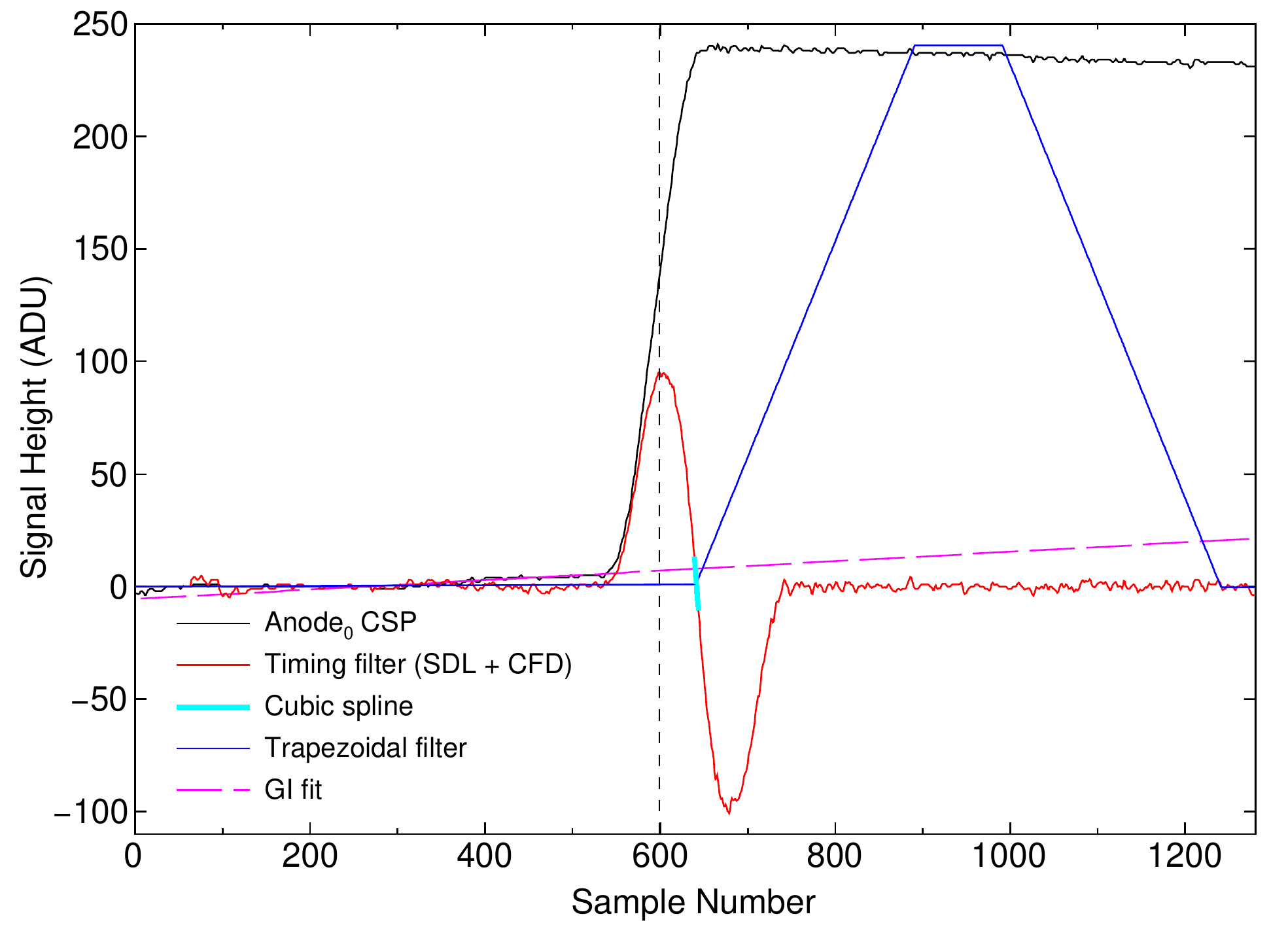}};
    \node[anchor=south east, inner sep=0, xshift=-0.15cm, yshift=0.6cm](wf) at (wf.center) {\includegraphics[trim={1.75cm 1.25cm 0.2cm 0.3cm}, clip, width=0.25\textwidth]{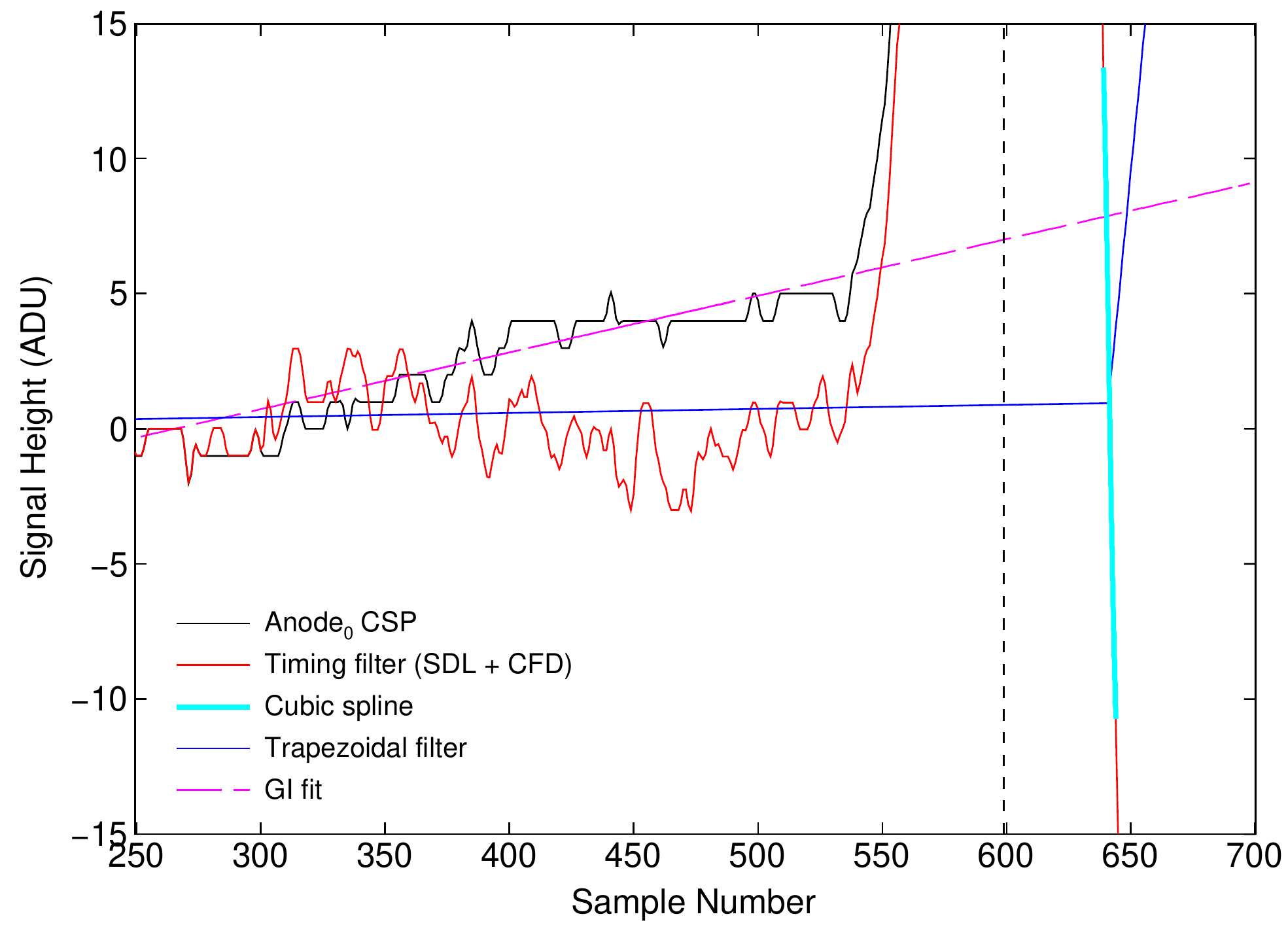}};
  \end{scope}
\end{tikzpicture}
\end{indented}
  \caption{\label{fig:waveforms} Digitized sample-side anode waveform (solid black curve) plotted with timing (solid red curve) and energy (solid blue curve) filters as well as the cubic spline interpolation of the timing filter zero crossing (thick, solid cyan curve) and linear fit used to calculate the grid inefficiency (GI - long broken magenta curve). The vertical dashed line marks the time at which the center of gravity of the electron cloud passes the Frisch-grid. The inset shows the detail of the zero crossing of the timing filter as well as the linear rise of the anode signal resulting from the GI.}
\end{figure}
The raw waveforms were analyzed using the analysis methods described in reference \cite{Mosby2014}, with a few key differences. The anode pulse heights were evaluated using a combination of finite impulse response filters during post-processing. First the trigger timing for the start of the energy filter was realized using a combination of a single delay line (SDL) filter (see \cite{Gerardi2014} and references therein) and a constant fraction discriminator (CFD). Next the trigger time was ascertained by interpolating the zero crossing of the timing filter with a cubic spline. Finally a trapezoidal filter identical to that described in reference \cite{Jordanov1994} was used to ascertain the anode pulse heights. A waveform from a typical fission event for the downstream anode along with energy and timing filters as well as the cubic spline used to interpolate the zero crossing is shown in \cref{fig:waveforms}. Pileup events (with $\alpha$ particles) were rejected broadly based on the ratio of their pulse height to pulse area.
\subsection{Neutron Time of flight}\label{sec:nTof}
\begin{figure}[t]
  \begin{tikzpicture}[scale=1]
    \begin{scope}
      \node[anchor=south west, inner sep=0, label={[align=left, xshift=-2.25cm, yshift=-0.6cm](a)}](nTof) at (0,0) {\includegraphics[trim={0.25cm 0 0 0.25cm}, clip, width=0.5\textwidth]{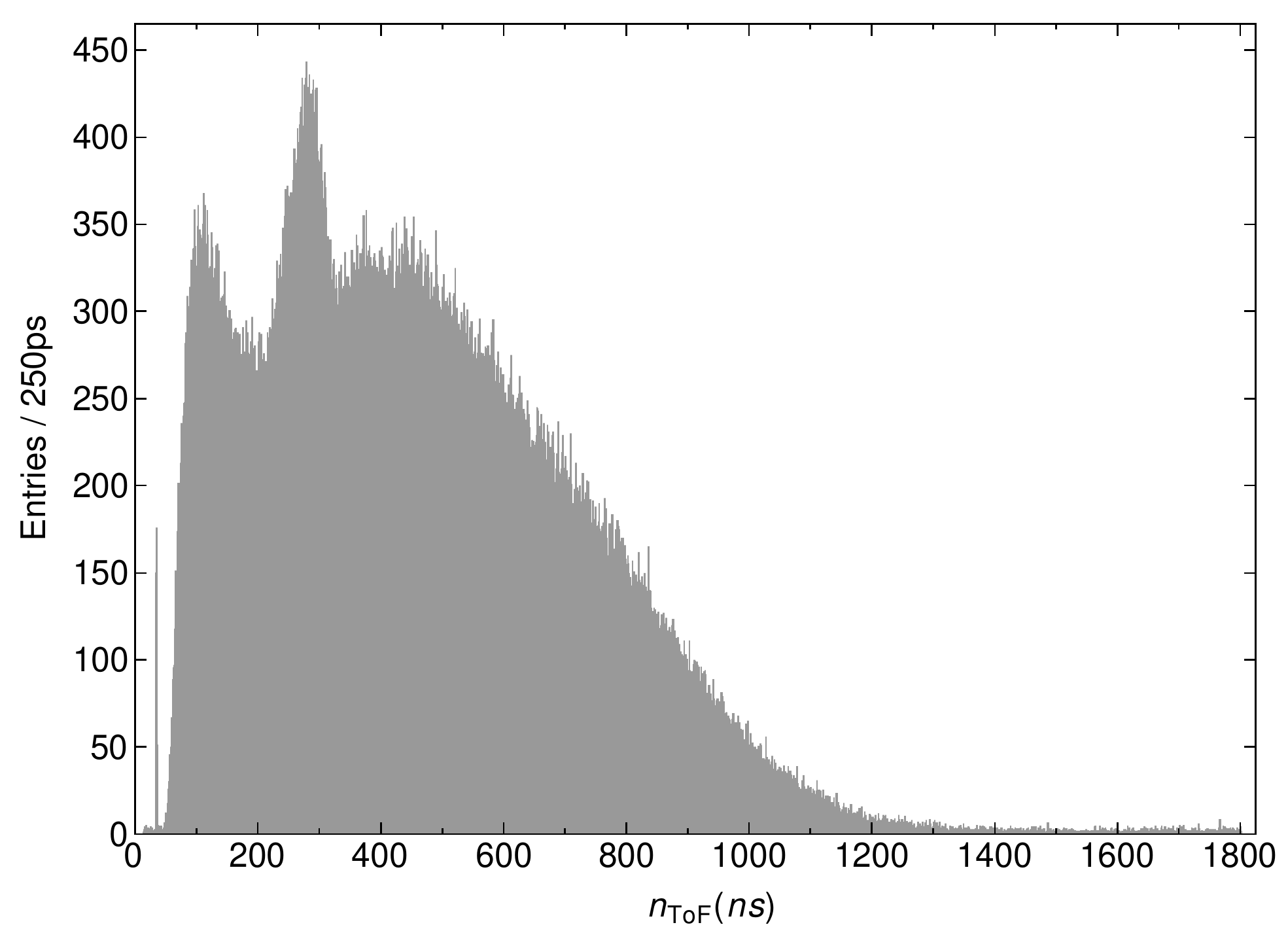}};
      \node[anchor=south west, inner sep=0, xshift=-0.4cm, yshift=-0.2cm] at (nTof.center) {\includegraphics[trim={1cm 0.75cm 0.25cm 0.2cm}, clip, width=0.25\textwidth]{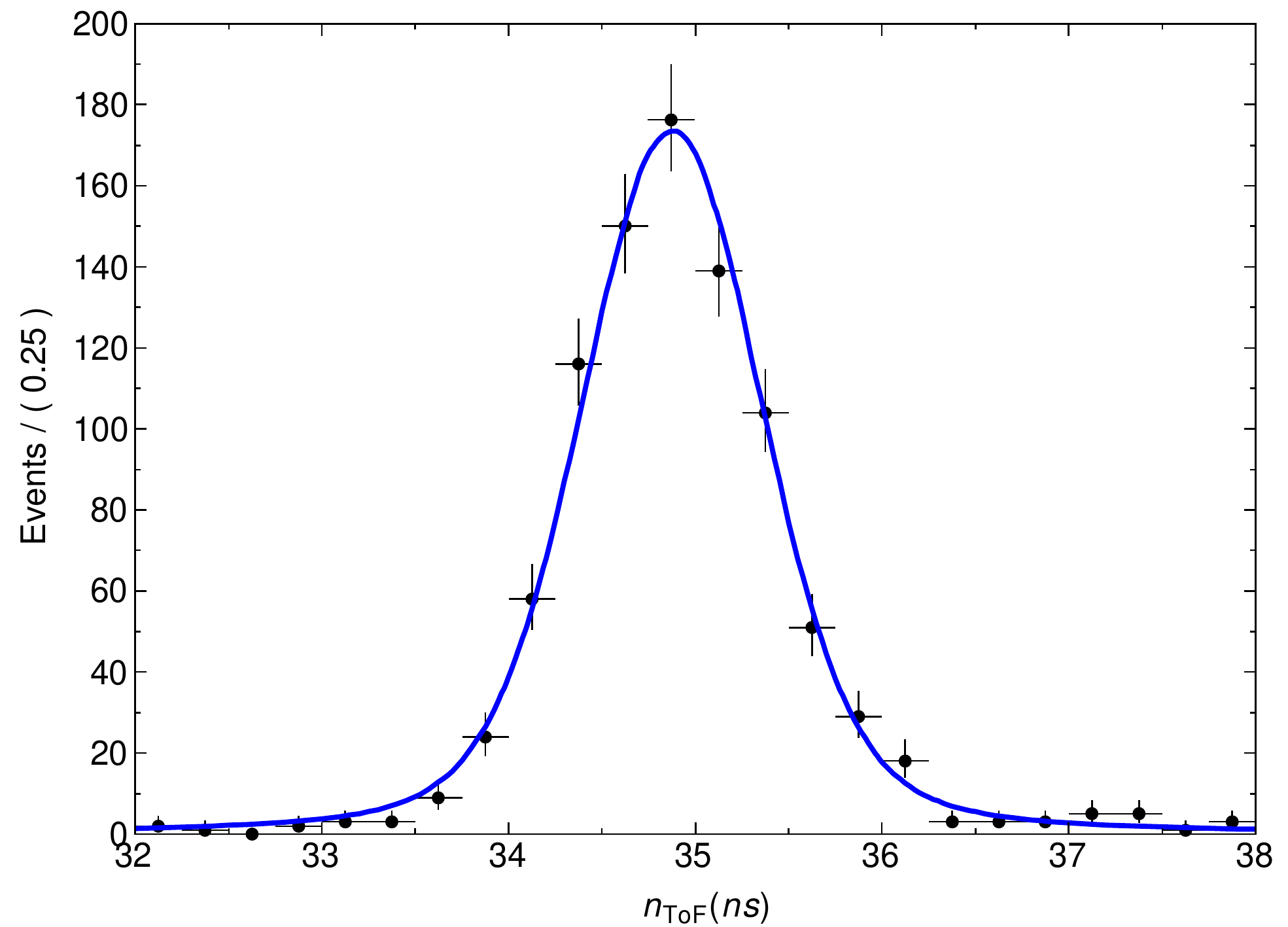}};
      \node[anchor=west, inner sep=0, label={[align=left, xshift=-2cm, yshift=-0.6cm](b)}] at (nTof.east) {\includegraphics[trim={0.25cm 0 0 0.25cm}, clip, width=0.5\textwidth]{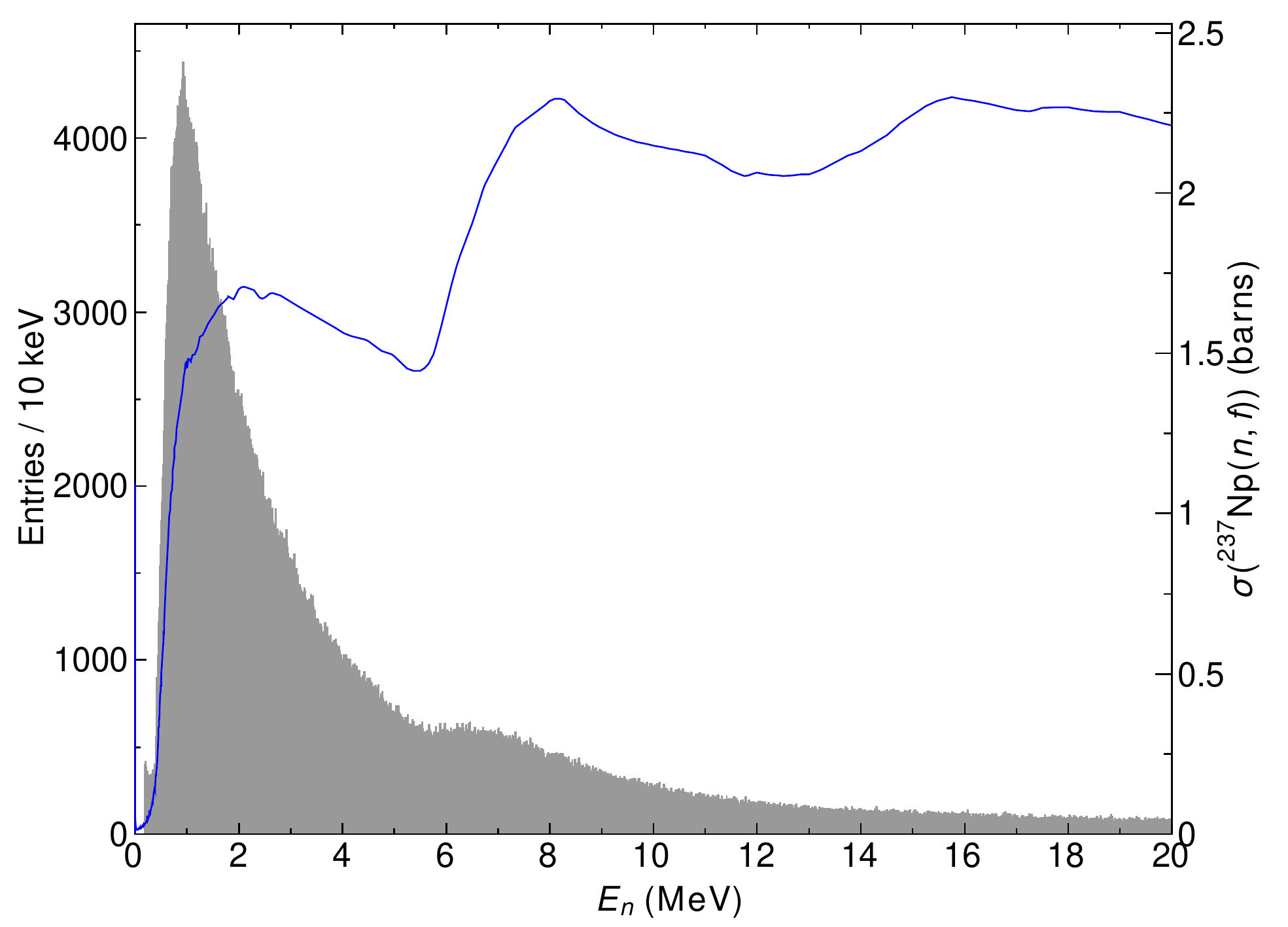}};
    \end{scope}
  \end{tikzpicture}
  \caption{\label{fig:nTof} (a) WNR neutron time of flight spectrum and photofission peak induced by the \ensuremath{\gamma}-flash from the W spallation target (inset). The photofission peak was fit with a Voigt profile in order to determine the FWHM timing resolution of the TFGIC (1.15~ns - see text for details). (b) Neutron energy spectrum obtained by applying \cref{eqn:En} to the neutron TOF spectrum plotted with the \npnf\ cross-section from the ENDF/B-VIII.0 evaluation \cite{Brown2018}.}
\end{figure}
The neutron time-of-flight (TOF) was measured by comparing the timing signal from the fast cathode to the the accelerator $T_0$ timing signal. The accelerator $T_0$ signal is a beam pickup detector located immediately upstream of the W spallation target. Using this signal as a reference to the timing signal of the cathode's fast preamplifier signal and knowledge of the pulse structure of the beam, one may reconstruct the neutron TOF. The neutron TOF spectrum for all fission events from the present measurement is shown in panel (a) of \cref{fig:nTof}. The inset of \cref{fig:nTof} shows the photofission peak - fission events induced by the \ensuremath{\gamma}-flash from the W spallation target. In order to determine the timing resolution for the TFGIC, the photofission peak was fit with a Gaussian convolved with a Lorentzian distribution (a.k.a.~the Voigt profile or Voigtian). The FWHM of the Voigtian fit was calculated using Eq.~(17) in reference \cite{AlOmar2020} to be 1.15 ns.\\
\indent The position of the \np{237} target (mounted in the center of the TFGIC) was measured to be a distance of $L = 10.62 \pm 0.01$~m from the W spallation target. This distance was measured by using the photofission peak in the neutron time of flight (TOF) spectrum. The neutron TOF spectrum was calibrated by placing a carbon filter in the neutron beamline between the detector and spallation target and utilizing characteristic features in the neutron TOF spectrum produced by $^{12}\mathrm{C}+n$ elastic scattering (which has a well known cross section) along with the photofission peak. Knowing the neutron TOF and flight path length, one may calculate the (relativistic) kinetic energy of an incident neutron as
\begin{equation}\label{eqn:En}
    E_n = m_nc^2\left(\frac{1}{\sqrt{1-\beta^2}} - 1\right)
\end{equation}
where $\beta = L \big/\Delta t_nc$, $L$ is the flight path length, $\Delta t_n$ is the neutron TOF, $c$ is the speed of light and $m_n$ is the neutron mass. The resulting neutron energy spectrum is shown in panel (b) of \cref{fig:nTof}. Because the WNR spallation target is a white source as well as the pulse structure of the beam, it is possible that low energy neutrons from previous beam pulses could arrive in coincidence with high energy neutrons from the current beam pulse. For the $10.62\pm0.01$~m flight path in this measurement, these so-called ``wrap around'' neutrons would have an energy of $E_n \le 182$~keV. At this energy, the \npnf\ cross section is 2 orders of magnitude lower than that of the requisite high energy neutrons they would come in coincidence with. Therefore, the wrap around neutrons have been deemed to have a negligible effect on the measurement and as such, they are ignored in the analysis of the present data. Additionally, the present analysis was restricted to neutrons with energies in the range $0.2~\mathrm{MeV}\le E_n\le 100~\mathrm{MeV}$.
\subsection{Anode pulse height corrections}\label{sec:aphCorr}
Differences between the two volumes of the TFGIC (such as upstream vs downstream and sample vs target side volume) cause small but measurable differences in the signal response of the anodes and grids of each respective volume. The following sections detail the methods employed to correct the anode pulse height spectra before performing further analysis. 
\subsubsection{Grid inefficiency}\label{sec:GI}
The grid inefficiency (GI) is a well-known phenomenon inherent in the response of Frisch-gridded ionization chambers \cite{Khriachkov1997, Al-Adili2012nima_a}. A small correction is therefore applied to the anode pulse height in order to correct for the GI. The value of the grid inefficiency can either be measured experimentally or calculated from the geometry of the ionization chamber using \cite{Bunemann1949}
\begin{equation}
    \eta_{\mathrm{GI}}(a, d, r) = \left(1+\frac{d}{\frac{a}{2\pi}\left[\left(\frac{\pi r}{a}\right)^2 - \ln\left(\frac{2\pi r}{a}\right)\right]}\right)^{-1}
\end{equation}
where $d$ is the distance between the anode and grid and $r$ and $a$ are the radius and pitch of the grid wires, respectively. From the dimensions stated in section \cref{sec:detector}, one calculates a value of $\eta_{\mathrm{GI}} = 0.028\pm 0.04$. Alternatively, the GI can be measured directly using the digitized anode waveforms (see inset of \cref{fig:waveforms} and reference \cite{Khriachkov1997}). The GI was determined experimentally using the method described in reference \cite{Khriachkov1997} to be $\eta_{\mathrm{GI}}^{\mathrm{exp}} = 0.027\pm0.03$, in good agreement with the above calculated value of $\eta_{\mathrm{GI}}$. The anode pulse heights were then corrected using the experimentally measured value and additive method described in references \cite{Al-Adili2012nima_a, Goeoek2012} via
\begin{equation}
    P_{\mathrm{a}}^{(\mathrm{Corr})} = \frac{P_{\mathrm{a}} - \eta_{\mathrm{GI}} P_{\mathrm{sum}}}{1 - \eta_{\mathrm{GI}}}
\end{equation}
where $P_{\mathrm{a}}$ is the anode pulse height extracted from the trapezoidal filter, and $P_{\mathrm{sum}} = P_{\mathrm{a}} + P_{\mathrm{g}}$ is the sum of the anode and grid pulse heights.
\subsubsection{Fission product polar angle}
\begin{figure}[t]
  \begin{tikzpicture}[scale=1]
    \begin{scope}
      \node[anchor=south west, inner sep=0, label={[align=left, xshift=-2.25cm, yshift=-0.75cm](a)}](xdcos) at (0,0) {\includegraphics[width=0.5\textwidth]{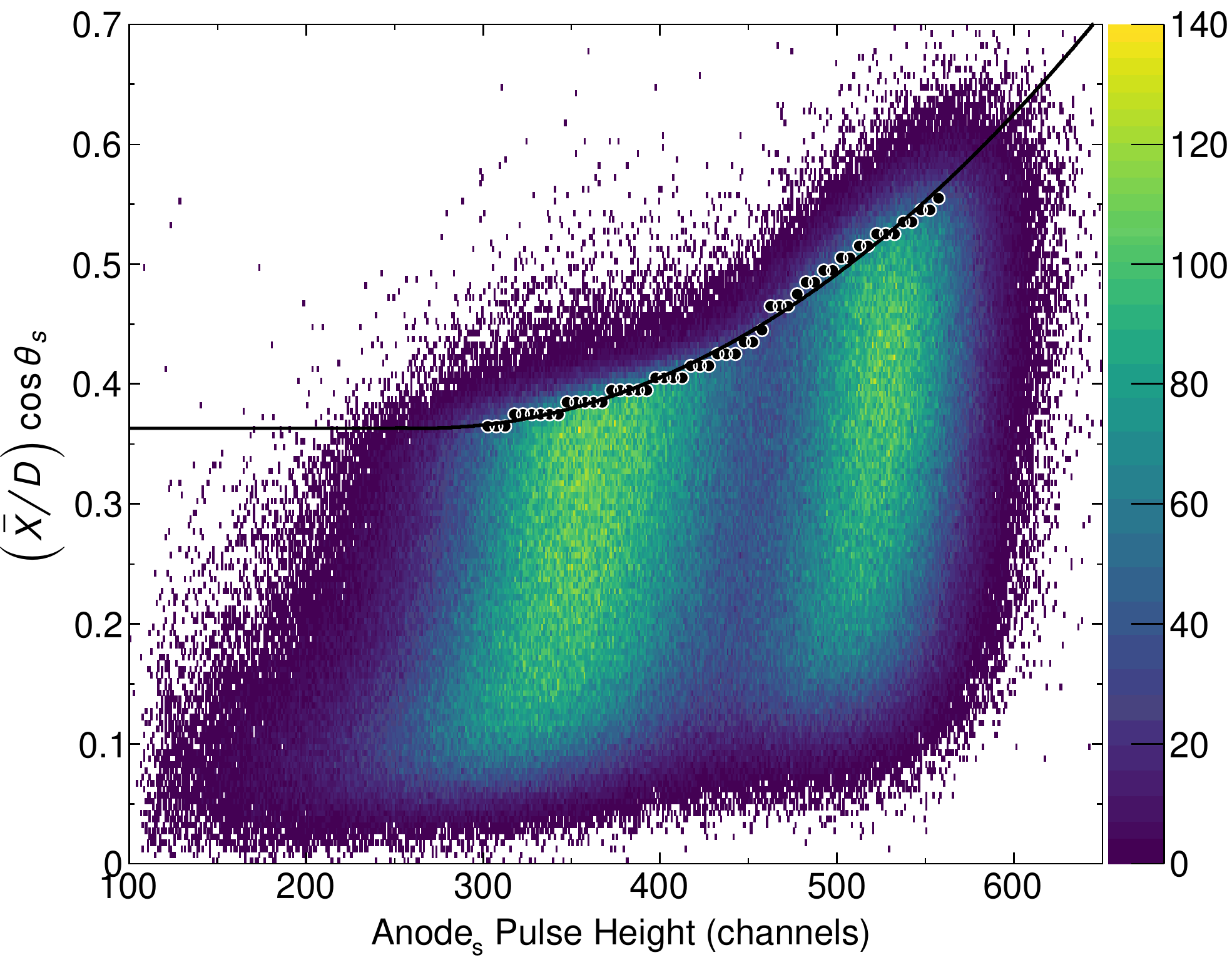}};
      \node[anchor=west, inner sep=0, label={[align=left, xshift=-2.875cm, yshift=-0.75cm](b)}] at (xdcos.east){\includegraphics[width=0.5\textwidth]{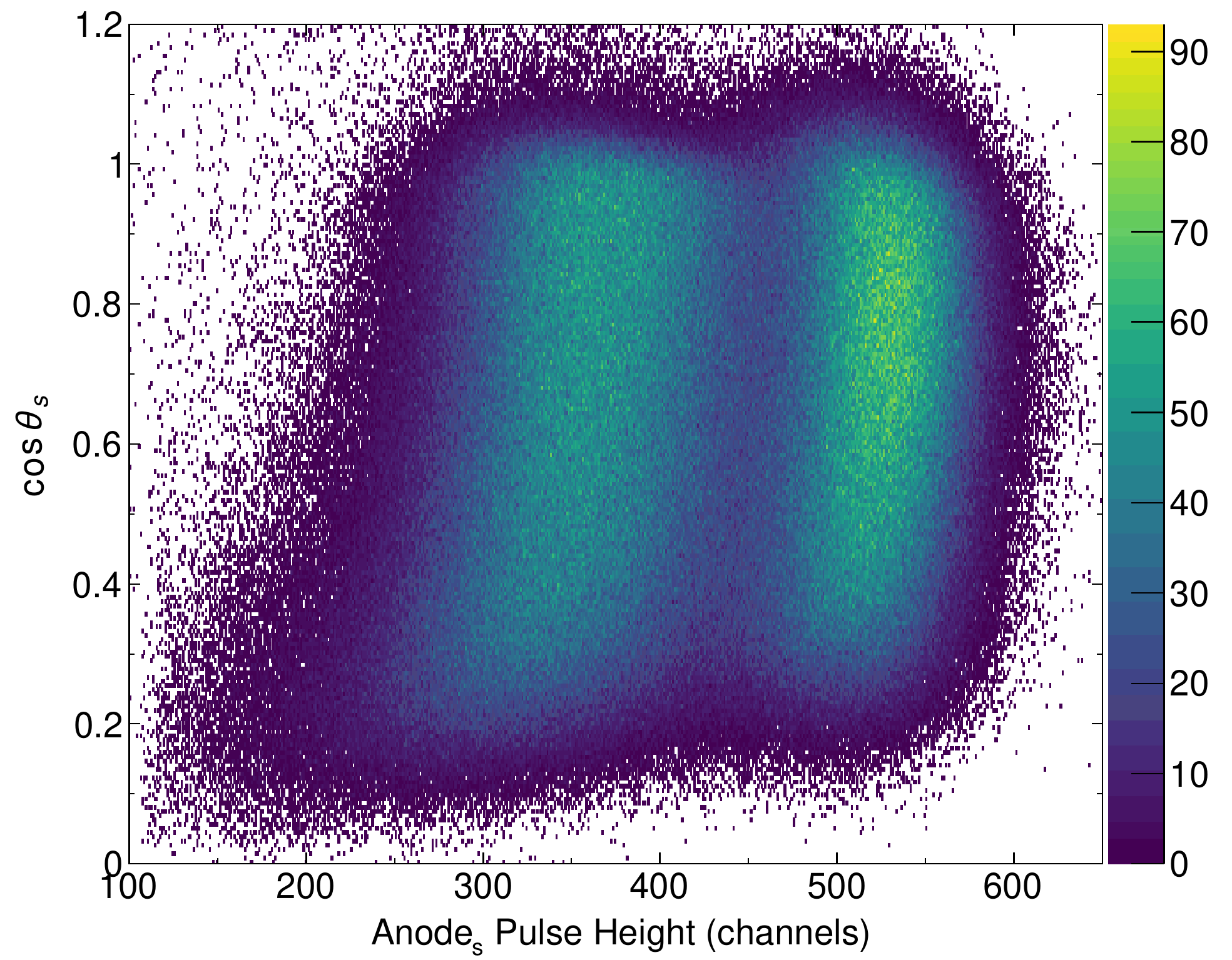}};
    \end{scope}
  \end{tikzpicture}
  \caption{\label{fig:xdcos_s} (a) 2-D distribution $\left(\overline{X}/D\right) \cos\theta$ vs anode pulse height for the sample side of the TFGIC. The edge of this distribution is found using a derivative filter and the edge points (black circles) are fit with a quadratic function (black solid curve) to correct the angular range. (b) Resulting corrected fragment angular distribution. See text for details.}
\end{figure}

The amount of charge collected on the grids of the TFGIC is proportional to the kinetic energy of the fission product as well as the polar angle $\theta$ of its ionization track. The ratio of the anode and grid pulse heights is given by \cite{Budtz-Joergensen1987}
\begin{equation}
    \frac{P_{\mathrm{g}}}{P_{\mathrm{a}}} = \frac{\overline{X}}{D}\cos\theta 
\end{equation}
where $\overline{X}$ is the distance from the point of fission to the center of mass (CM) of the electron cloud and $D$ is the distance between the cathode and grids. The 2-D distribution resulting from plotting $\left(\overline{X}/D\right) \cos\theta$ vs anode pulse height for the sample side is shown in \cref{fig:xdcos_s}. Because the angular distribution of fission products is isotropic, the angular distribution should fill the range $\cos\theta = [0, 1]$. The value of $\overline{X} /D$ for a given fission product depends on mass, charge and energy, hence the varying range of fission products as a function of anode pulse height. In order to obtain the angular distribution of fission products the edge of the distribution in \cref{fig:xdcos_s} is found using a derivative filter and the edge points are fit with a quadratic function in order to ascertain the maximum value $\overline{X}/D$ and normalize the distribution to this range.
\subsubsection{Incident neutron momentum}\label{sec:nMom}
The incident neutrons impart a small amount of momentum to the compound nucleus of the fissioning system. Therefore, a transformation to the center of mass (CM) frame is necessary in order to accurately compare the signals from each half of the chamber. In order to avoid ambiguities in the mass of the compound nucleus resulting from multichance fission, this transformation is applied provisionally for events with incident neutron energies of $E_n \leq 1.0\,\mathrm{MeV}$ and then applied to the data set as a whole after additional anode corrections are made. The CM transformation is made using the following (nonrelativistic) relation
\begin{equation}
    E^{\mathrm{CM}}_{i} = E^{\mathrm{lab}}_{i} \pm 2\,\mathcal{M}_i \sqrt{E^{\mathrm{lab}}_{i} E^{\mathrm{lab}}_{n}}\cos\theta^{\mathrm{lab}}_{i} + \mathcal{M}_i^2 E^{\mathrm{lab}}_n
\end{equation}
where $\mathcal{M}_i = \sqrt{\mu_im_n/M^{2}_\mathrm{CN} }$, $\mu_{0,1} = E^{\mathrm{lab}}_{1,0}M_{\mathrm{CN}}\big/(E_0+E_1)$, $E^{\mathrm{lab}}_n$ is the lab frame incident neutron energy, $E^{\mathrm{lab}}_i$ is the lab frame energy of the FF in the corresponding volume ($i=0,1$ for the upstream and downstream volumes, respectively), $m_n$ is the neutron mass and $M_{\mathrm{CN}}$ is the mass of the $^{237}\mathrm{Np}+n$ compound nucleus.
\subsubsection{Energy loss in the target and backing}\label{sec:e-loss}
Fission fragments lose a fraction of their kinetic energy as they travel through the target and backing material. The energy loss of fission products traveling through matter is dependent on the product's proton number $Z$, energy $E$, the stopping power of the medium through which the fission product is traveling $dE/dx$, and the distance traveled through that material $x$. Because the sample and backing material are of differing compositions and thicknesses, fission products lose a different amount of energy depending on which side of the target they exit. For $2E$ experiments using Si detectors, the energy loss of fission products through the target material is often calculated using models and/or tabulated values for the stopping powers for a given medium and ion species combination (see e.g.~reference \cite{Pica2020}), which introduces yet another source of systematic uncertainty to the measurement. On the other hand, with the TFGIC, the energy loss of fission products is determined empirically from the data. Because the distance a fission product travels through the target/backing depends on the polar angle of its trajectory, so too does the energy loss of that fission product. \\
\begin{figure}[t]
\begin{tikzpicture}[scale=1]
  \begin{scope}
    \node[anchor=south west, inner sep=0, label={[align=left, xshift=-3cm, yshift=-0.65cm](a)}](eLoss) at (0,0) {\includegraphics[trim={0.25cm 0 0.375cm 0.25cm}, clip, width=0.625\textwidth]{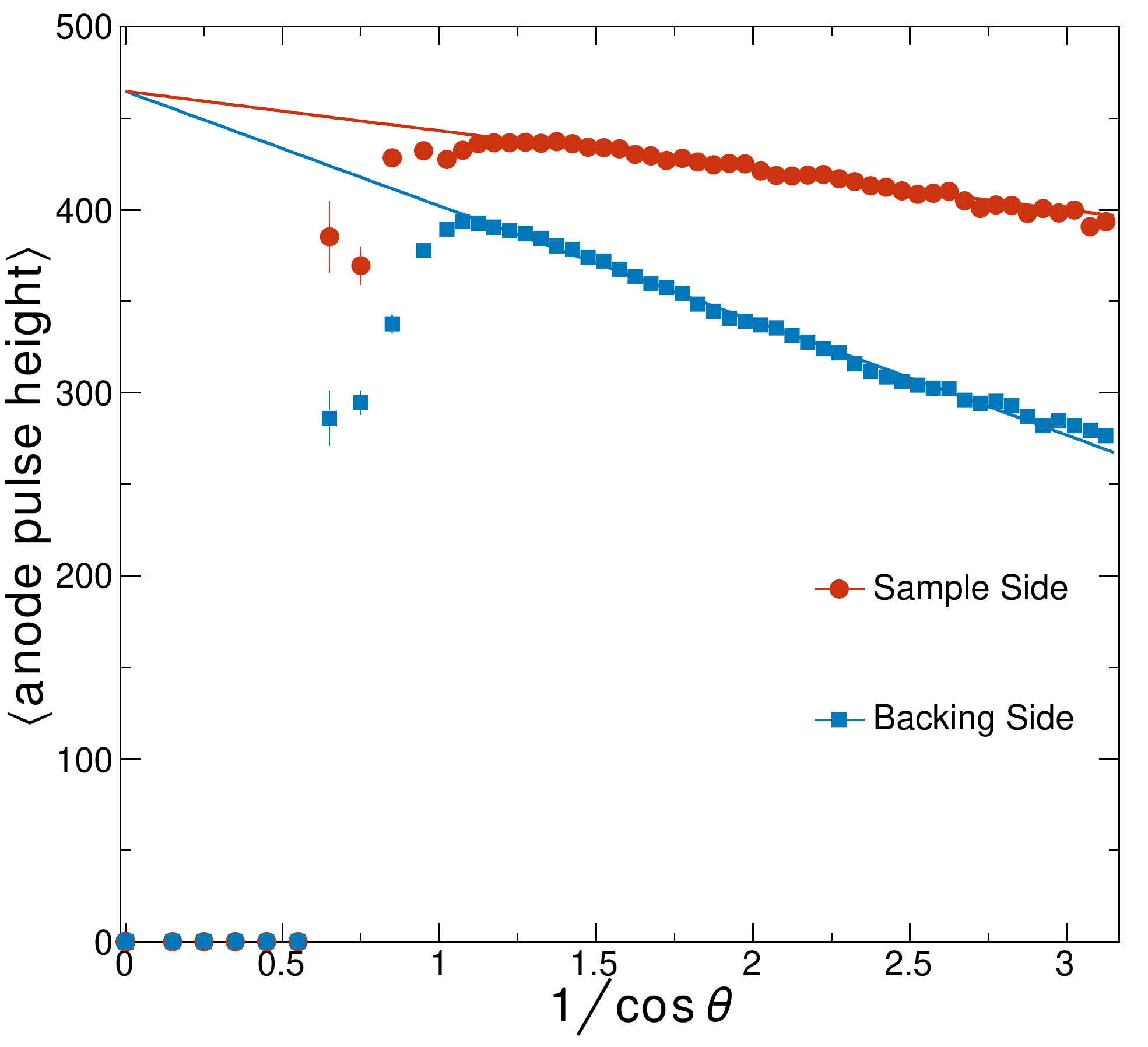}};
    \node[anchor=west, inner sep=0, xshift=0.cm, yshift=0.cm, label={[align=left, xshift=-2.25cm, yshift=-0.4cm](b)}](elSchem) at (eLoss.east) {\includegraphics[trim={0cm 0cm 0.2cm 0cm}, clip, width=0.375\textwidth]{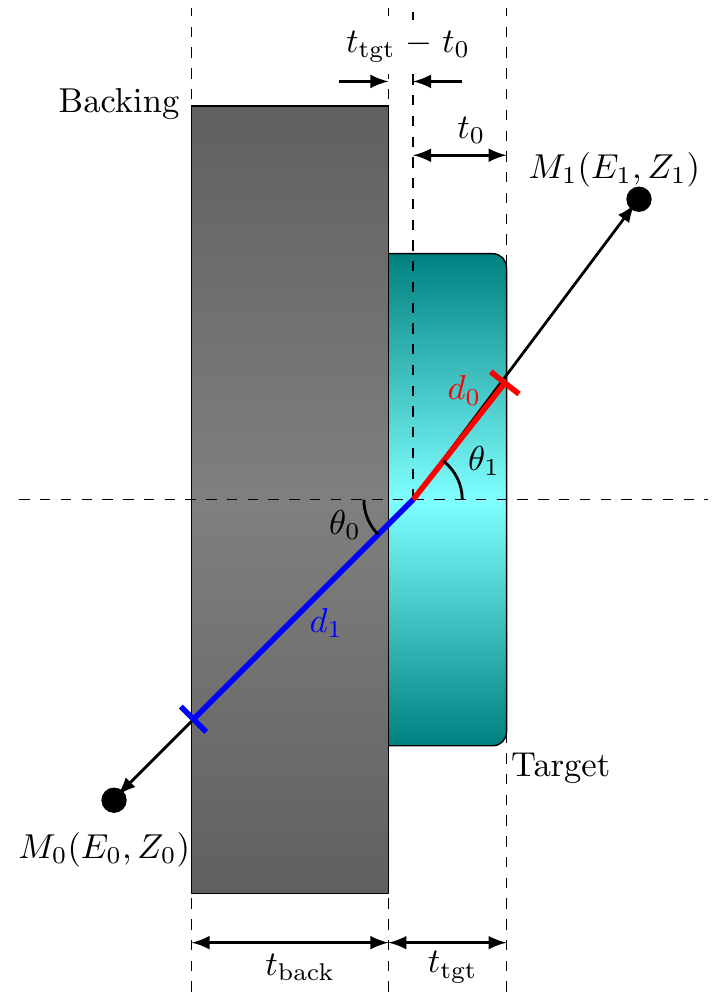}};
  \end{scope}
\end{tikzpicture}
  \caption{\label{fig:eLoss} Panel (a) - Average anode pulse height vs $1\big/\cos\theta_{0,1}$ for the target (red circles) and backing (blue squares) volumes. Panel (b) - Schematic illustrating the distances travelled by fission products through the target and backing.}
\end{figure}
\indent Consider the special case of an infinitesimally thin target and backing. The measured anode pulse heights in both the target and backing volumes in this case ($P_{\mathrm{ideal}}$) would simply be proportional to the kinetic energies of the fission products (with the constant of proportionality depending on the response of the TFGIC). Since the target and backing are of finite extent, the measured anode pulse heights are reduced by the energy loss of the fission products through the target and backing materials. From panel (b) of \cref{fig:eLoss}, it is clear that the measured anode pulse height is then
\begin{equation}
    P_{\mathrm{tgt}}^{\mathrm{meas.}} = P_{\mathrm{ideal}} - \frac{\Delta s_{\mathrm{tgt}}t_0}{\cos\theta_1}\label{eqn:tEloss}
\end{equation}
where $\Delta s_{\mathrm{tgt}}$ is the stopping power of the fission product $M_1$ in the target material. Similarly for the backing side, one finds
\begin{equation}
    P_{\mathrm{back}}^{\mathrm{meas.}} = P_{\mathrm{ideal}} - \frac{\Delta s_{\mathrm{back}}t_{\mathrm{back}} + \Delta s_{\mathrm{tgt}}(t_{\mathrm{tgt}} - t_0) }{\cos\theta_0}\label{eqn:bEloss}
\end{equation}
where $\Delta s_{\mathrm{back}}$ is the stopping power of the fission product $M_0$ in the C backing. This is an intractable problem for individual fission products, given that the stopping power of both the target and backing depend sensitively on the mass, charge and energy of the ion species traveling through it, and it is impossible to know the depth within the target at which fission occurred. Therefore averages of these quantities are used to correct the anode pulse height scales. Taking the averages of \cref{eqn:tEloss,eqn:bEloss} one sees they have the form 
\begin{equation}
    \langle P_{\mathrm{back}}^{\mathrm{meas.}}\rangle = \langle P_{\mathrm{ideal}}\rangle - \frac{m_{\Delta s}}{\cos\theta_0}
\end{equation}
and one may ascertain the energy loss correction factor for the anode pulse height scales with a linear fit to the data in \cref{fig:eLoss} where $m_{\Delta s}$ and $\langle P_{\mathrm{ideal}}\rangle$ are free parameters of the fit.
\begin{figure}[t] 
\begin{indented} 
\item[]
  \includegraphics[width=0.5\textwidth]{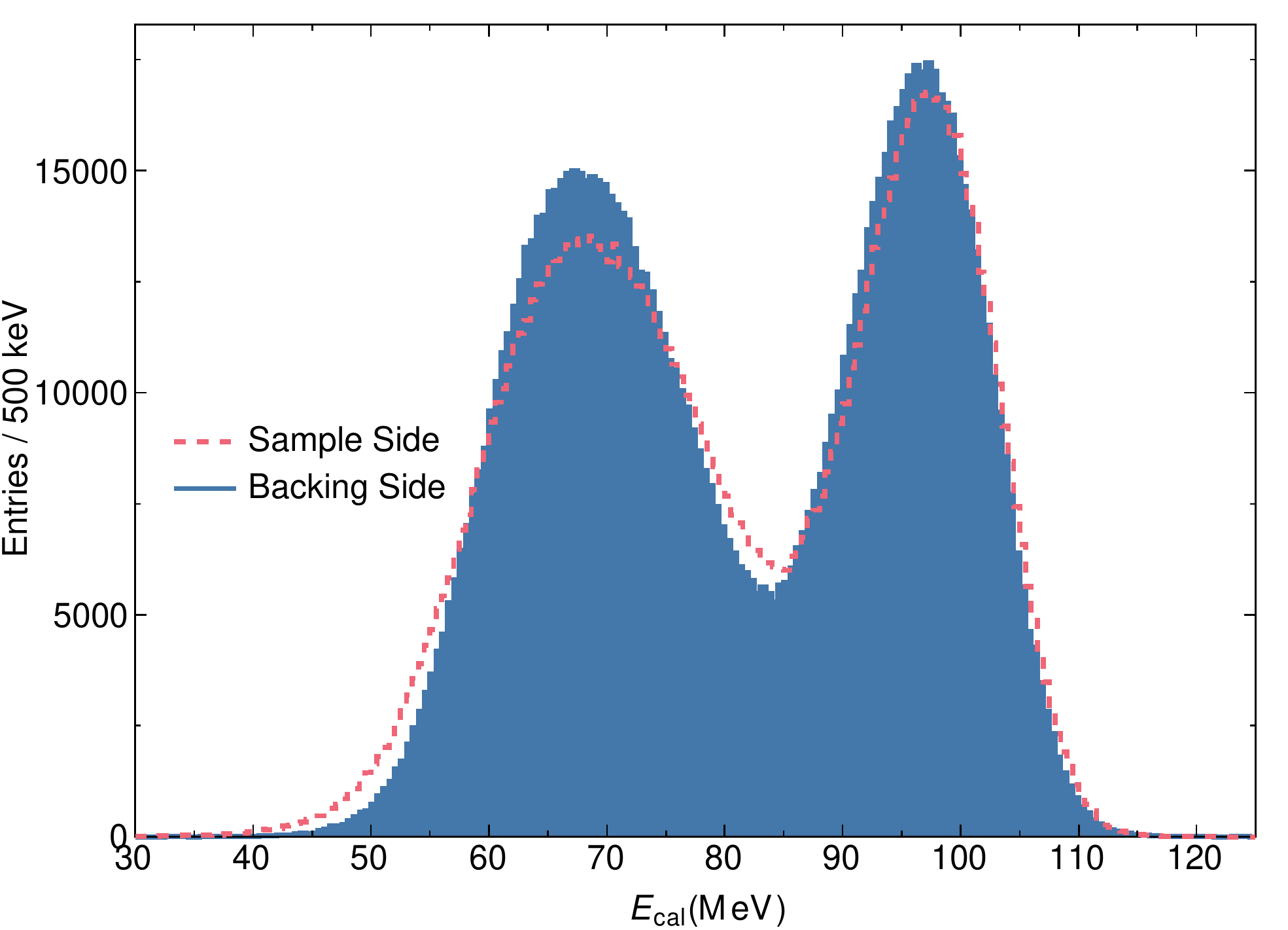}
  \end{indented}
  \caption{\label{fig:correctedAnodes} Corrected and energy calibrated anode pulse height spectra for both the sample side anode (red histogram) and backing side anode (blue histogram).}
\end{figure}
\subsubsection{Preamplifier gain correction}\label{sec:gainCorr}
The final correction to the anode pulse height spectra is a small correction due to the slight difference between the gains of each anode's respective preamplifier. The energy loss-corrected pulse height spectra of each volume was fitted with a double-Gaussian distribution in order to precisely locate the heavy and light mass peaks. A linear correction was then applied to the backing side pulse height spectrum. Once these corrections were made, the data were selected via a cut on polar angle ($\cos\Theta_{\mathrm{CM}} \ge 0.5$) and an upper limit on incident neutron energy for calibration and further physics analysis.
\subsection{Energy calibration}\label{sec:eCal}
\begin{table}[b]
\caption{Parameters used for energy calibration of the anode pulse height spectra. See text for details. Fission product mass and velocity values are from reference \cite{Naqvi1986}.\label{tab:eCal}}
\begin{indented}
\item[] 
\begin{tabular*}{0.55\textwidth}{@{\extracolsep{\fill}} llr}
   \br
   Parameter                        & Light             & Heavy \\
   \midrule
   $\left<m_i\right>$ (u)           & $97.07\pm0.06$    & $138.21\pm 0.06$  \\
   $\left<v_i\right>$ (cm/ns)       & $1.3997\pm0.0013$ & $0.9874\pm0.0010$ \\
   $\left<E_i\right>$ (MeV)         & $98.55\pm0.06$      & $69.83\pm0.03$        \\
   $\left<\Delta E_i\right>$ (MeV)  & $1.59\pm0.11$      & $2.60\pm0.11$      \\
   $PHD_i$ (MeV)                    & $3.67\pm0.003$    & $4.43\pm0.001$    \\
   $E_i^{\text(cal)}$ (MeV)         & $96.47\pm0.13$        & $68.00\pm0.11$        \\
   \br
 \end{tabular*}
 \end{indented}
\end{table}
Energy calibration of the anode pulse height spectra was achieved using the method described in references \cite{Higgins2018, Higgins2020}. The centroids of the light and heavy peaks in each anode pulse height spectrum for a subset of events with $E_n\leq1$~MeV were normalized to the absolute values of $\left<E_i\right>$ reported by Naqvi \etal\ at an incident neutron energy of $E_n = 0.8$~MeV \cite{Naqvi1986} reduced by the requisite PHD of the TFGIC for the mean post-neutron evaporation mass. The relevant parameters used for energy calibration are presented in \cref{tab:eCal}.\\
\indent Using equation (3) in reference \cite{Naqvi1986} along with the reported average measured velocities of light and heavy fragments ($v_0 = 1.3997\pm0.0013$~cm/ns and $v_1 = 0.9874\pm0.0010$~cm/ns, respectively), one finds $\langle TKE\rangle_{\mathrm{pre}} = 0.5M_{\mathrm{CN}}v_0v_1 = 170.49$~MeV. This is 4.31~MeV less than the reported value of $\langle TKE\rangle_{\mathrm{pre}} = 174.80$~MeV \cite{Naqvi1986}. Similarly, using the average light fragment mass $\langle m^*_0\rangle = 98.66\pm0.06$~u reported by Naqvi \etal\ along with the aforementioned light fragment velocity, one finds $\langle E_0^* \rangle = 100.17$~MeV, which is 1.61~MeV lower than the value of $\langle E^*_0\rangle = 101.78$~MeV reported in section III A of reference \cite{Naqvi1986}. Therefore, it is assumed that the velocity values reported by Naqvi \etal\ in Table II of reference \cite{Naqvi1986} are not corrected for energy loss of the fission products through the target material and backing as well as the SSB dead layer. Using the relationships for pre- and post-neutron evaporation energies (\cref{eqn:Epost-pre}), one would expect post-neutron evaporation energies of $\langle E_0\rangle = 100.14$~MeV and $\langle E_1\rangle = 72.43$~MeV for the light and heavy products, respectively. These values are respectively $1.59$~MeV and $2.60$~MeV greater than the energies of $\langle E_{0}\rangle = 98.55$~MeV and $\langle E_1\rangle = 69.83$~MeV found using the values of mean fragment velocities and \cref{eqn:KE}. Thus, we correct $\langle E_i\rangle$ in \cref{tab:eCal} using the assumed energy loss values $\langle\Delta E_i\rangle$ (also given in \cref{tab:eCal}), where the stated uncertainties on $\langle\Delta E_i\rangle$ are those derived in reference \cite{Muller1981}.\\
\indent The values for PHD$_i$ were obtained using the experimental values of the PHD for the TFGIC obtained by Vives \cite{Vives1998} interpolated with a cubic spline for the requisite $\left<m_i\right>$. Subtracting the respective PHD$_i$ from $\left<E_i\right>$ yields the final energy calibration parameters $E_i^{(\text{cal})}$. Equating these values to the centroids of the light and heavy peaks in the anode pulse height spectra gives a linear energy calibration of the pulse height spectra scale. The resulting corrected and energy calibrated pulse height spectra for both the sample side and backing side anodes is shown in \cref{fig:correctedAnodes}.
\subsection{The 2E method}\label{sec:2E}
Fission is a complex process and therefore many factors (such as prompt neutron evaporation, multichance fission, pre-equilibrium pre-scission phenomena) need to be taken into consideration in order to determine \tke\ and fission mass distributions as a function of incident neutron energy. The $2E$ method employs conservation of momentum and nucleon number to correct for prompt neutron evaporation and multichance fission to calculate the primary fission fragments' masses based on their measured kinetic energies. The following sections describe in detail the $2E$ method as well as the treatment of fission phenomena employed in the present work.
\subsubsection{Mass determination\label{sec:massCalc}}
Pre-neutron evaporation fragment masses are calculated on an event-by-event basis using an iterative process. For a given event, the incident neutron energy $E_n$ and the measured kinetic energies of the fission products $E_i$ are used as inputs for the iterative loop. In order to treat multichance fission, the compound nuclear mass is determined at this point by selecting a random mass from compound nuclear mass probability distributions obtained from Monte-Carlo simulations of the de-excitation of fission fragments based on the incident neutron energy (see section \cref{sec:multiChance}). The iterative loop is then initiated beginning with the ansatz
\begin{equation}\label{eqn:ansatz}
    m_{i,j = 0}^* = \frac{M_{\mathrm{CN}}}{2} 
\end{equation}
where $M_{\mathrm{CN}}$ is the mass of the compound nucleus, $i = {0,1}$ and $j$ is the iteration index. After scission, conservation of momentum and nucleon number demand
\begin{align}
     m_0^*E_0^* =&\;m_1^*E_1^* \label{eqn:mom-E} \\
    m_0^* + m_1^* =&\;M_{\mathrm{CN}} \label{eqn:mcn}
\end{align}
Inserting \cref{eqn:mom-E} into \cref{eqn:mcn} and rearranging we have
\begin{equation}
    m_{0,1}^* = \frac{E_{1,0}^*M_{\mathrm{CN}}}{E_0^*+E_1^*}\label{eqn:mPost}
\end{equation}
\begin{figure*}[t]
  \includegraphics[clip, width=\textwidth]{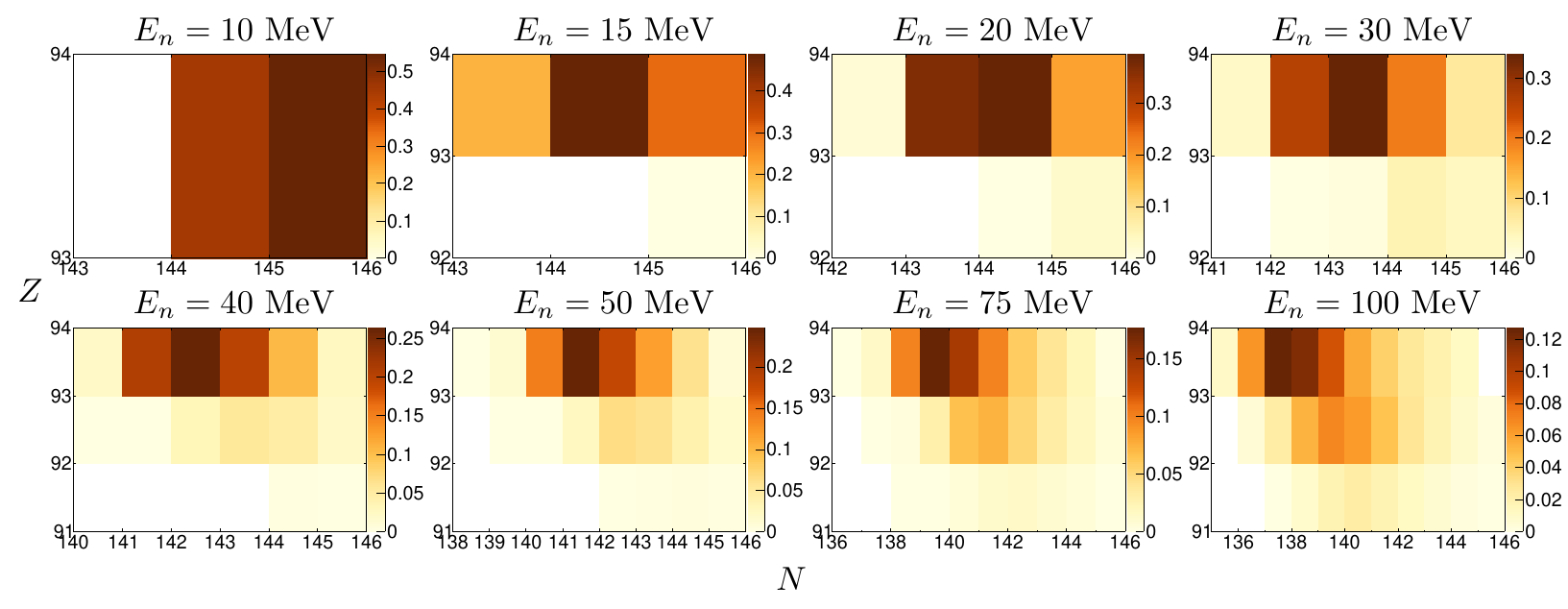}
  \caption{\label{fig:GEFMcnProb} Compound nuclear mass distributions ($Z$ vs $N$) for several incident neutron energies from the GEF simulation of \npnf\ \cite{Schmidt2016, Jurado2015}.}
\end{figure*}
where $m_i^*$ are the pre-neutron evaporation fragment masses and $E_i^*$ are the kinetic energies of the pre-neutron evaporation fragments. Assuming that the prompt neutrons are emitted isotropically by fully accelerated fragments, then on average the fragment velocity is unchanged by the neutron emission and the post-neutron evaporation kinetic energies are related to the pre-neutron evaporation kinetic energies by
\begin{equation}
    E_i = \frac{m_i}{m_i^*}E_i^*\label{eqn:Epost-pre}
\end{equation}
The post-neutron evaporation masses are
\begin{equation}
    m_{i,j} = m_{i,j}^* - m_n\nu(m_{i,j}^*, E_n)\label{eqn:Mpost-pre}
\end{equation}
where $m_n$ is the mass of the neutron and $\nu(m_{i, j}^*, E_n)$ is the prompt neutron multiplicity (which is in general dependant on the primary fragment mass and the incident neutron energy). Since it is impractical to measure the actual number of neutrons emitted by each fragment, a statistical approach using the average prompt neutron multiplicity $\langle\nu\rangle(m_i^*, E_n)$ to calculate primary fragment masses with \cref{eqn:Mpost-pre} is employed. However, data on $\langle\nu\rangle(m_i^*, E_n)$ is sparse \cite{Talou2021}, so a Monte Carlo simulation of the de-excitation of fission fragments is used to determine $\langle\nu\rangle(m_i^*, E_n)$ (see section \cref{sec:nubar}). Once the post-neutron evaporation masses are calculated, the measured kinetic energies of the fission products can be corrected for PHD using the measured PHD as a function of ion mass (see section \cref{sec:phd}). Using \cref{eqn:Mpost-pre}, we can rewrite \cref{eqn:Epost-pre} as
\begin{equation}
    E_i^* = E_i\left(1 + \frac{\langle\nu\rangle(m_i^*, E_n)}{m_i}\right)\label{eqn:Epre-post}
\end{equation}
It is now convenient to define
\begin{equation}
    \zeta \equiv \frac{m_0m_1^*}{m_0^*m_1} =  \frac{1+\langle\nu\rangle_1\big/m_1}{1+\langle\nu\rangle_0\big/m_0}\label{eqn:zeta}
\end{equation}
where we have defined $\langle\nu\rangle_i \equiv \langle\nu\rangle(m_{i,j}^*, E_n)$. Inserting \cref{eqn:Epre-post} and \cref{eqn:zeta} into \cref{eqn:mPost} gives
\begin{align}
        m_{0, j+1}^* &= \frac{M_{\mathrm{CN}}}{1+E_0\big/(E_1\zeta)}\label{eqn:mPre0} \\
        m_{1, j+1}^* &= \frac{M_{\mathrm{CN}}}{1 + E_1\zeta\big/E_0}\label{eqn:mPre1}
\end{align}
Once the pre-neutron evaporation fragment masses have been calculated with \cref{eqn:mPre0,eqn:mPre1}, they are compared to the pre-neutron evaporation fragment masses from the previous iteration. If the convergence condition $\abs{m_{i,j} - m_{i, j+1}} \le 0.05$ is met, the pre-neutron evaporation fragment energies are then calculated using \cref{eqn:Epost-pre} and the next event is processed. Otherwise, the $j+1$ values are saved and the loop is repeated until the convergence condition is met or a set number of iterations is surpassed and the event is discarded.
\begin{figure}[t]
  \begin{tikzpicture}[scale=1]
    \begin{scope}
      \node[anchor=south west, inner sep=0, label={[align=left, xshift=2.5cm, yshift=-0.6cm](a)}](nubar) at (0,0) {\includegraphics[trim={0.3125cm 0.06125cm 0.18125cm 0.25cm}, clip, width=0.5\textwidth]{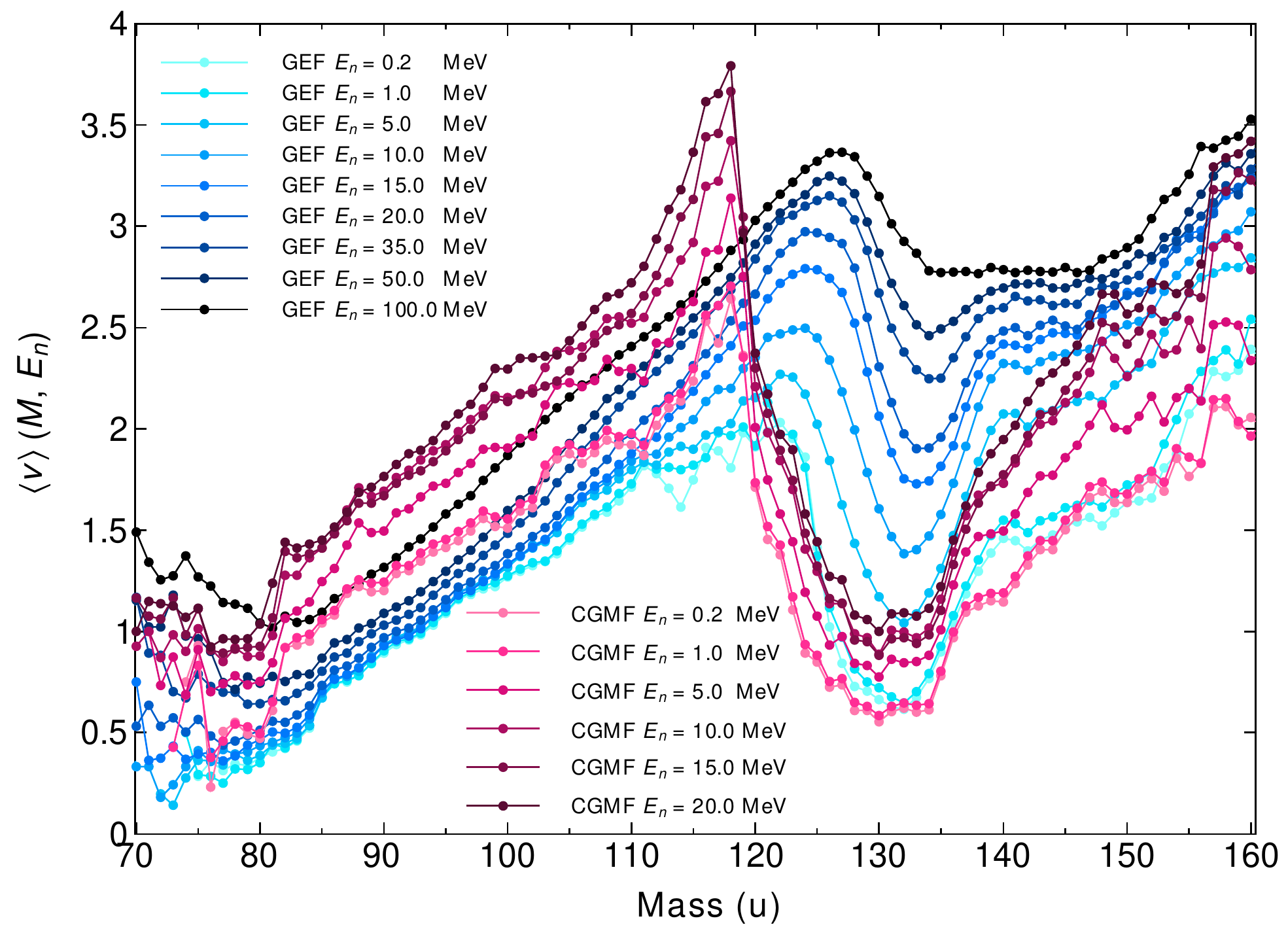}};
      \node[anchor=west, inner sep=0, label={[align=left, xshift=2.5cm, yshift=-0.6cm](b)}] at (nubar.east) {\includegraphics[trim={0.3125cm 0.06125cm 0.25cm 0.25cm}, clip, width=0.5\textwidth]{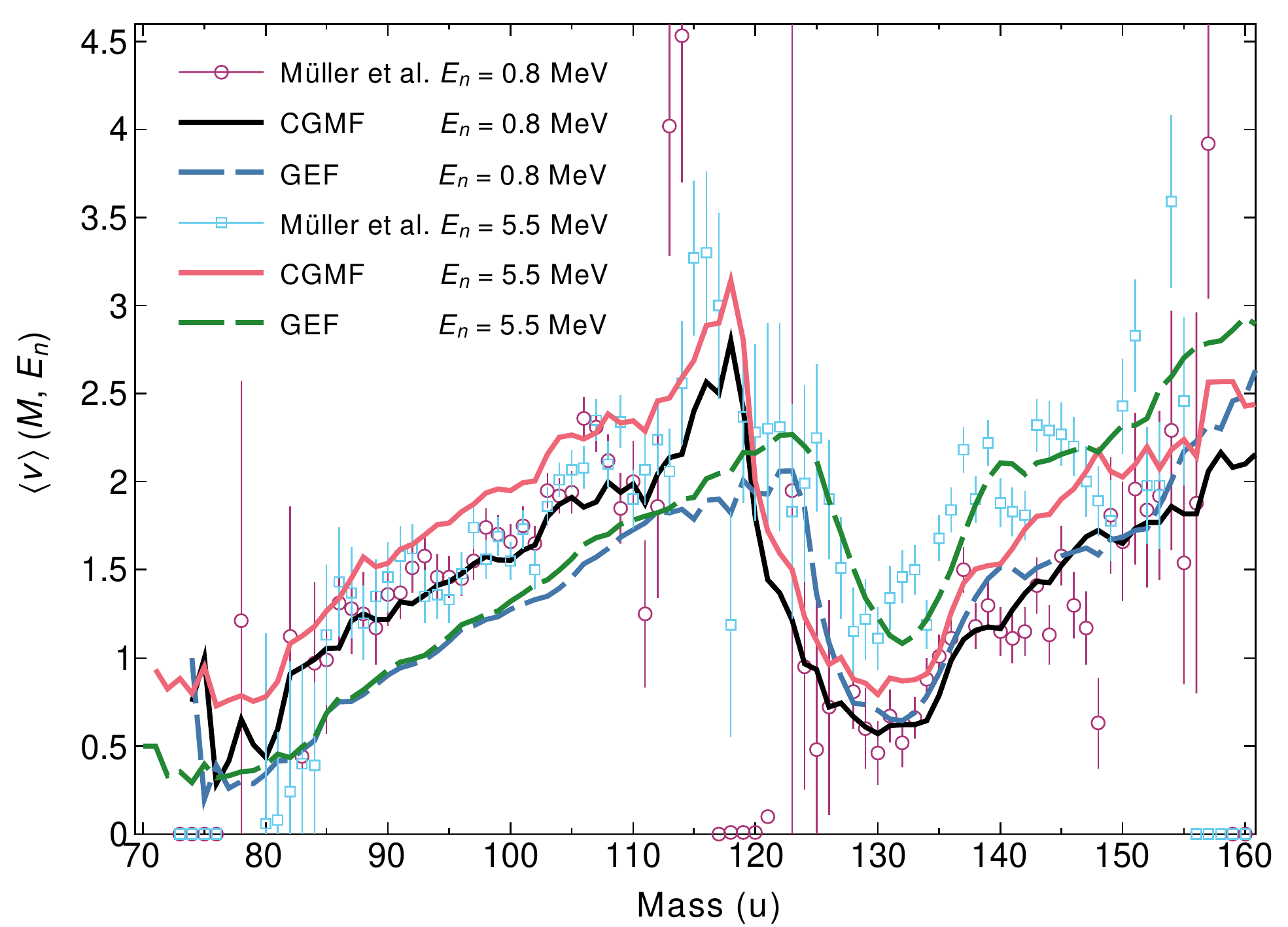}};
    \end{scope}
  \end{tikzpicture} 
  \caption{\label{fig:nubarVsA} (a) Average neutron multiplicity as a function of primary fragment mass for several incident neutron energies $E_n$ obtained from GEF\cite{Schmidt2016, Jurado2015} and CGMF\cite{Talou2021, cgmf} model simulations. (b) Average neutron multiplicity as a function of primary fragment mass data from reference \cite{Muller1981} at incident neutron energies of $E_n = 0.8$~MeV and $E_n = 5.5$~MeV compared to neutron sawtooths obtained from GEF \cite{Schmidt2016, Jurado2015} and CGMF \cite{Talou2021, cgmf} simulations at the same energies.}
\end{figure}
\subsubsection{Multichance fission}\label{sec:multiChance}
Multichance fission occurs when one or more neutrons and/or protons are emitted from the compound nucleus before scission occurs. The probability of different modes of multichance fission occuring increases with increasing excitation energy of the compound nucleus (and hence, increasing incident neutron energy). In order to treat multichance fission in this work, the General Description of Fission Observables (GEF) \cite{Schmidt2016, Jurado2015} and Cascade Gamma Multiplicities from Fission (CGMF) \cite{Talou2021} codes were used to generate compound nuclear mass probability distributions at the incident neutron energies listed in section \cref{sec:nubar}. Multichance fission was then treated event-by-event when calculating post-neutron evaporation product masses via the $2E$ method by randomly selecting a mass from the compound nuclear mass distribution for the appropriate incident neutron energy range and using that mass as the initial condition (\cref{eqn:ansatz}) for the $2E$ iterative loop. Compound nuclear mass probability distributions obtained from GEF for several incident neutron energies are shown in \cref{fig:GEFMcnProb}.
\subsubsection{Prompt neutron evaporation\label{sec:nubar}}
Since the $2E$ method only measures the energies of correlated fission products, some assumptions must be made about prompt neutron evaporation from the fission fragments. This can be done a number of different ways (c.f. - \cite{Hambsch2000, Duke2016}). Contemporarily, the most commonly employed method utilizes the average neutron multiplicity as a function of primary fragment mass and incident neutron energy (so-called neutron sawtooths) obtained from Monte-Carlo simulations of the decay / de-excitation of fission fragments after scission to deduce pre- and post-neutron evaporation masses from the measured correlated fission product energies. It has been demonstrated that the pre- and post-neutron evaporation \tke\ is only weakly influenced by prompt neutron emission \cite{Hambsch2000}. On the other hand, the determination of pre- and post-neutron evaporation masses is strongly dependent on prompt neutron emission.\\
\begin{figure*}[t]
    \begin{indented} 
    \item[]
        \includegraphics[trim={0.375cm 0.125cm 0.49cm 0.25cm},clip,width=0.8\textwidth]{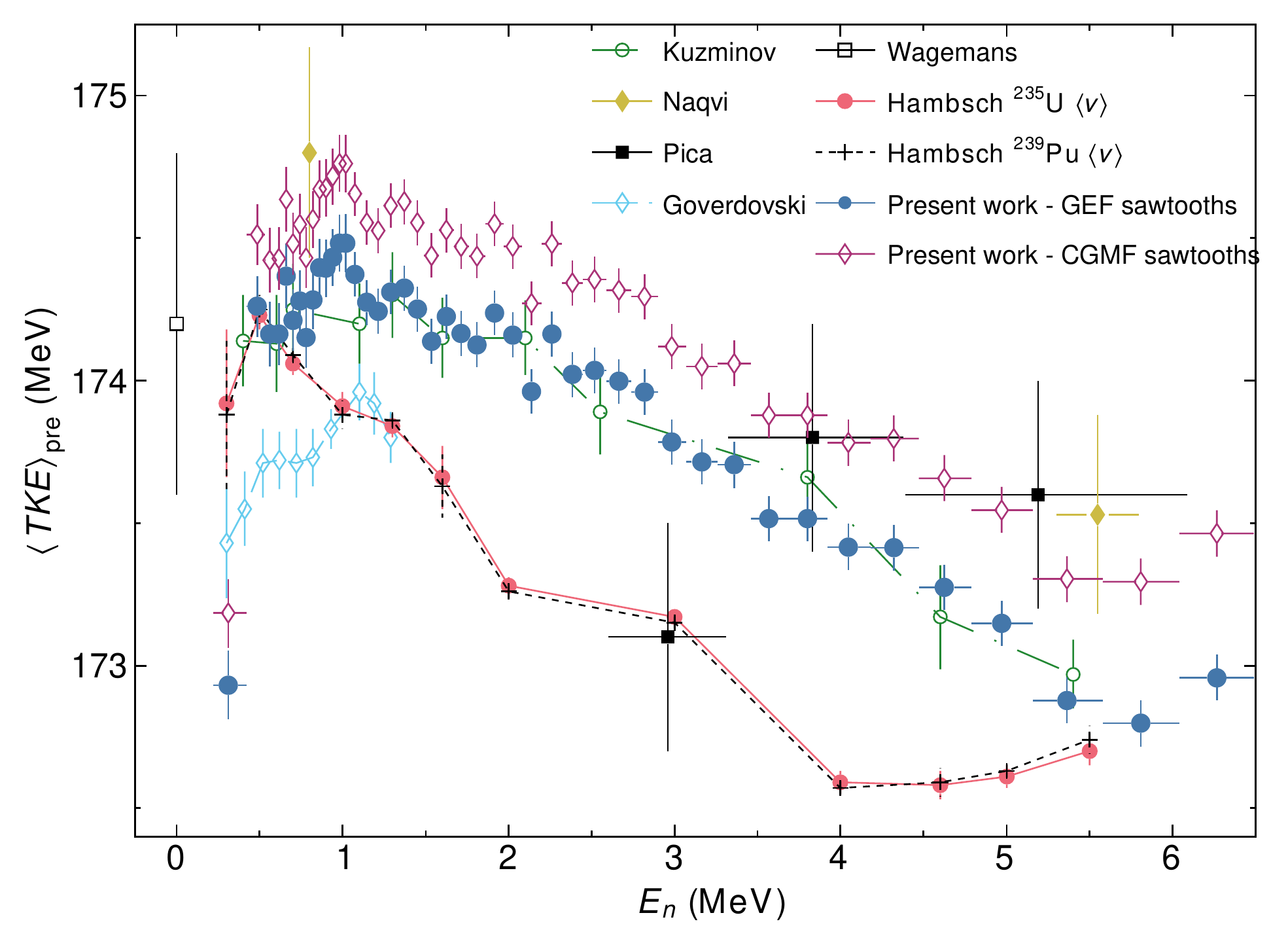}
    \end{indented}
    \caption{\label{fig:np237le} Present data compared to existing pre-neutron evaporation \npnf\ \tke\ vs $E_{n}$ data
\cite{Kuzminov1970, Wagemans1981, Naqvi1986, Goverdovski1992, Hambsch2000, Pica2020}. Data from references \cite{Bennett1967, Asghar1977, Ruiz1978} excluded from this plot. Data from Kuzminov \cite{Kuzminov1970}, Goverdovski \cite{Goverdovski1992} and Wagemans \cite{Wagemans1981} have been renormalized using the currently accepted value of $\langle TKE\rangle_{\mathrm{pre}}$ for thermal neutron induced fission of ${}^{235}$U. See text for details.}
\end{figure*}
\indent For the present analysis, neutron sawtooths for incident neutron energies from $E_n = 0.2 - 20.0$~MeV were obtained from two different codes of the Monte-Carlo simulation of the de-excitation of fission fragments - GEF \cite{Schmidt2016, Jurado2015} and CGMF \cite{Talou2021}. Presently, the CGMF model only performs simulations of \npnf\ for incident neutron energies up to $E_n = 20.0$~MeV, so only GEF neutron sawtooths and compound nuclear mass probability distributions (see section \cref{sec:multiChance}) were used in the the $2E$ analysis for $E_n > 20.0$~MeV. Neutron sawtooths obtained from both the GEF and CGMF simulations are plotted for various incident neutron energies in \cref{fig:nubarVsA}. The CGMF and GEF model codes were run for $5\times10^5$ events at the following incident neutron energies: $E_n = 2.5\times10^{-8}$~MeV (thermal), $E_n = 0.1 - 1.0$~MeV in $0.1$~MeV steps, $E_n = 1.0 - 10.0$~MeV in $0.5$~MeV steps, and $E_n = 10.0 - 20.0$~MeV in $1.0$~MeV steps. Additionally, the GEF simulations were run for $5\times10^5$ events for the incident neutron energies of $E_n = 20.0 - 100.0$~MeV in $5.0$~MeV steps. Neutron sawtooths obtained from the CGMF and GEF simulations at $E_n = 0.8$~MeV and $E_n = 5.5$~MeV are compared to the measured \nubar\ from M\"{u}ller \etal\ \cite{Muller1981} in panel (b) of \cref{fig:nubarVsA}.
\subsubsection{Fragment pulse height defect\label{sec:phd}}
The pulse height defect (PHD) for fission fragments was handled on an event-by-event basis during the $2E$ mass calculation (see section \ref{sec:massCalc}). Once the post-neutron evaporation product masses were calculated, the PHD was interpolated from the measured PHD as a function of ion mass that was measured by Vives \etal\ for an identical chamber \cite{Vives1998}. 
\subsection{Uncertainty Budget}
Statistical uncertainties on \tke\ and mass yields varied as a function of the incident neutron energies. For the energy range $0.2 \le E_n < 1.05$~MeV and the energy range $E_n \ge 10.12$~MeV, the neutron energy binning was selected such that a statistical uncertainty of $\sigma_{\langle TKE\rangle} \le 1.0\%$ was achieved. In the energy range $1.05 \le E_n < 10.12$~MeV, the neutron energy binning was selected such that $\sigma_{\langle TKE\rangle} \le 0.7\%$ was achieved. In addition to these statistical uncertainties, there are several sources of systematic uncertainty present in the measurement. These uncertainties are presented in \cref{tab:uncertainty} and include the following sources of systematic uncertainties: uncertainty due to the finite timing resolution of the fast cathode $\left(\sigma_{n\mathrm{TOF}} = 1.6\%\right)$, uncertainty due to the energy resolution of the TFGIC $\left(\sigma_{E} = 0.5\%\right)$, uncertainty due to the energy calibration procedure $\left(\sigma_{E_{\mathrm{Cal}}} = 0.2\%\right)$ and uncertainty due to the measurement of the PHD of the TFGIC $\left(\sigma_{\mathrm{PHD}} = 0.1\%\right)$. \\
\begin{table}[t]
\caption{\label{tab:uncertainty}Sources of systematic uncertainty in \tke\ and FPY resulting from measurement uncertainties.}
\begin{indented}
\item[]
\begin{tabular*}{0.5\textwidth}{@{\extracolsep{\fill}} llr}
   \br
   Source of Uncertainty        & Symbol & Value(\%)\\
   \midrule
   Neutron ToF & $\sigma_{n\mathrm{TOF}}$ & $1.6$\\
   TFGIC Energy Resolution & $\sigma_{E}$ & $0.5$ \\
   Energy Calibration & $\sigma_{E_{\mathrm{Cal}}}$ & $0.2$ \\
   Pulse Height Defect & $\sigma_{\mathrm{PHD}}$ & $0.1$ \\
   \br
 \end{tabular*}
 \end{indented}
\end{table}
\indent Another source of systematic uncertainty that contributes to the uncertainty in the measured mass yields is that associated with the simulation used for the de-excitation of fission fragments (see \cref{sec:nubar}). This value is difficult to quantify and depends sensitively on parametric uncertainties of the simulation. The analysis required to quantify this uncertainty is beyond the scope of this paper and hence is not considered further. 
\section{Results \label{sec:Results}}
A total of approximately $1.53\times10^6$ fission events were recorded over the three week irradiation period. Of these, approximately $1.15\times10^6$ fission events were selected for physics analysis via cuts described in \cref{sec:Analysis}. The results of the $2E$ analysis (described above) are presented in the following sections. 
\subsection{Average total kinetic energy of fission products}
\begin{figure}[t]
    \begin{tikzpicture}[scale=1]
    \begin{scope}
      \node[anchor=south west, inner sep=0, label={[align=left, xshift=2.75cm, yshift=-0.75cm](a)}](tke) at (0,0) {\includegraphics[trim={0 0.125cm 0 0.25cm}, clip, width=0.5\textwidth]{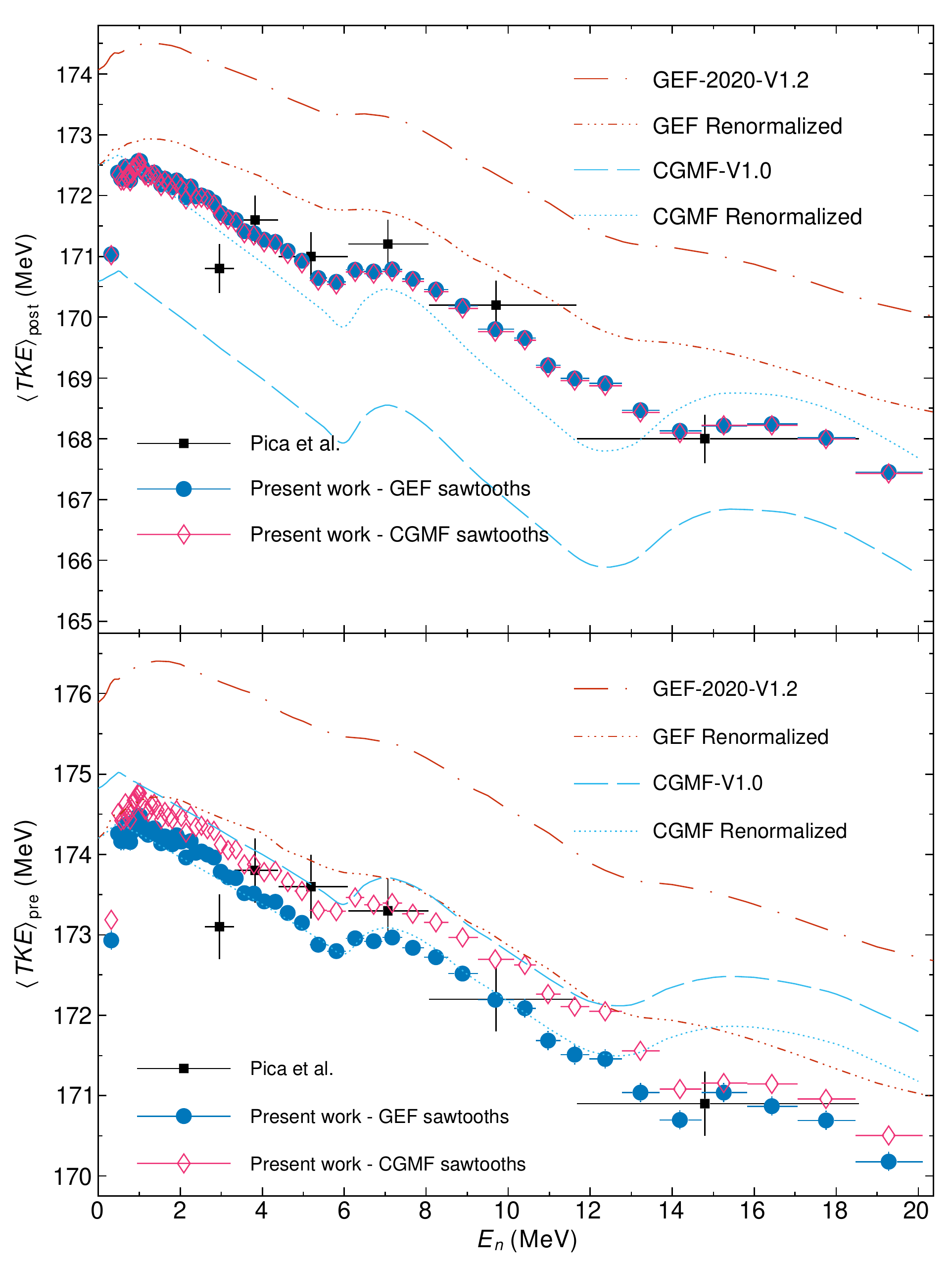}};
      \node[anchor=west, inner sep=0, label={[align=left, xshift=2.75cm, yshift=-0.75cm](c)}](tke2) at (tke.east) {\includegraphics[trim={0 0.125cm 0 0.25cm}, clip, width=0.5\textwidth]{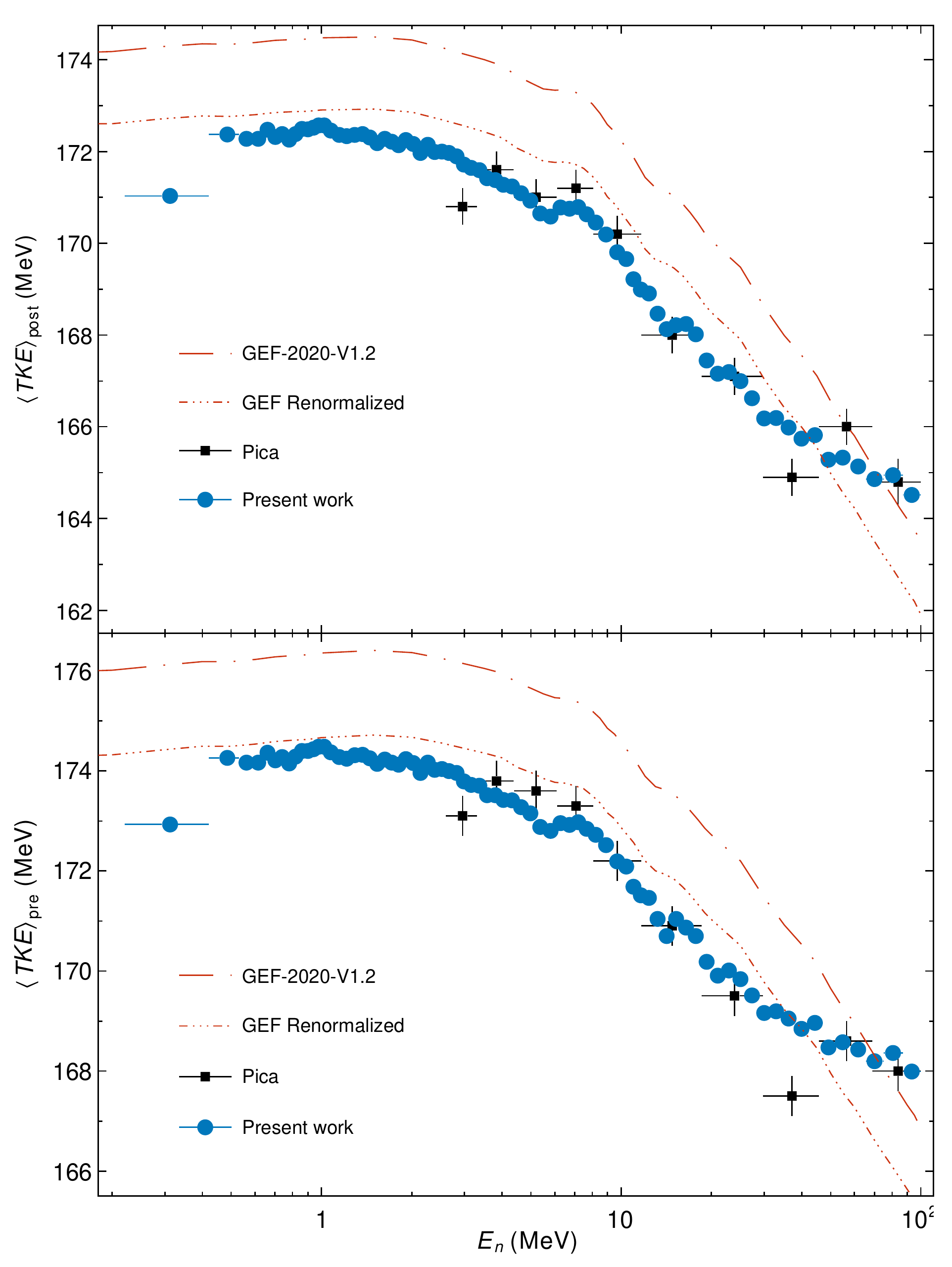}};
      \node[anchor=center, inner sep=0, label={[align=left, xshift=2.75cm, yshift=-0.75cm](b)}] at (tke.center) {};
      \node[anchor=west, inner sep=0, label={[align=left, xshift=2.75cm, yshift=-0.75cm](d)}] at (tke2.center) {};
    \end{scope}
    \end{tikzpicture}
    \caption{\label{fig:tkeVsEn} ((a) and (c)) Pre- and ((b) and (d)) Post-neutron evaporation \tke\ vs $E_n$ for the present work using neutron sawtooths from GEF \cite{Schmidt2016, Jurado2015} (blue filled circles) and CGMF \cite{Talou2021, cgmf} (magenta open diamonds) up to $E_n = 20$~MeV ((a) and (b)) and up to $E_n = 100$~MeV ((c) and (d)) using the GEF neutron sawtooths. The data are compared to the GEF (red dot-dashed curves) and CGMF \cite{Talou2021} (cyan long broken curves) model predictions, as well as the reference \cite{Pica2020} data (black open squares). The cyan dashed curves (in (a) and (b)) and the red triple-dot dashed curves (all panels) are respectively the result of renormalizing the GEF and CGMF models to the corrected \npnthf\ \tke\ value stated in \cref{sec:Wagemans}.}
\end{figure}
Pre-neutron evaporation \tke\ results for incident neutron energies of $E_n \le 6.6$~MeV using neutron sawtooths from the GEF and CGMF models in the $2E$ analysis are presented in \cref{fig:np237le} and compared to previous data. Pre- and post-neutron evaporation \tke\ results for incident neutron energies in the range $0.0 \le E_n \le 20.0$~MeV using both the GEF and CGMF neutron sawtooths in the $2E$ mass calculation are displayed in panels (a) and (b) of \cref{fig:tkeVsEn}. Pre- and post-neutron evaporation \tke\ results for incident neutron energies in the range $0.0 \le E_n \le 100.0$~MeV using GEF neutron sawtooths in the $2E$ mass calculation are displayed in panels (c) and (d) of \cref{fig:tkeVsEn} and compared to the results of reference \cite{Pica2020}. The red dot-dashed curves in \cref{fig:tkeVsEn} are the raw prediction of the GEF \cite{Schmidt2016, Jurado2015} model and the cyan dashed curves are the CGMF \cite{Talou2021,cgmf} predictions. For thermal neutron induced fission of \np{237}, GEF and CGMF respectively predict values of $\tke_{\mathrm{pre}} = 175.9$~MeV and $\tke_{\mathrm{pre}} = 174.82$~MeV, which are (respectively) $1.7$~MeV and $0.62$~MeV higher than the renormalized value of $\tke_{\mathrm{pre}} = 174.2\pm0.6$~MeV stated in section \cref{sec:Wagemans}. Renormalizing the GEF prediction to the renormalized value for \npnthf\ yields the triple-dot-dashed red curves in \cref{fig:tkeVsEn}. Similarly, renormalizing the CGMF prediction to the renormalized value for \npnthf\ yields the dotted cyan curves in \cref{fig:tkeVsEn}. The \tke\ data used to generate \cref{fig:tkeVsEn} are presented in \cref{tab:tke}.\\
\indent Examining \cref{fig:tkeVsEn}, one observes a significant increase in the \tke\ values of the experimental data relative to the GEF model's prediction beginning at incident neutron energies of approximately $E_n \approx 20-30$~MeV. This phenomenon has been noted in several previous fission studies utilizing the $2E$ method and performed at the WNR facility \cite{King2017, Yanez2018, Chemey2020, Pica2020}. However, the interpretation of this observation in the data is unclear. Although the $2E$ method employed in this study does account for multichance fission, it does not account for pre-equilibrium pre-fission neutron evaporation, which has been observed in the $^{239}\mathrm{Pu}(n,f)$ prompt fission neutron spectrum (PFNS) at incident neutron energies as low as $E_n \approx 15$~MeV \cite{Kelly2018}. Neither does the $2E$ analysis in this and previous studies account for the possibility of incomplete linear momentum transfer, which has been observed in Neutron Induced Fission Fragment Tracking Experiment (NIFFTE) \cite{Heffner2014a} data and is described in reference \cite{Hensle2020}. These effects could significantly impact and possibly bias the measurement at these incident neutron energies. Additional and more careful investigations at these energies are necessary to draw a definitive conclusion from the observed discrepancy.
\begin{figure*}[t]
  \begin{tikzpicture}[scale=1]
    \begin{scope}
      \node[anchor=south west, inner sep=0](yield) at (0,0) {\includegraphics[trim={0.35cm 0.05cm 0.375cm 0.2cm}, clip, width=\textwidth]{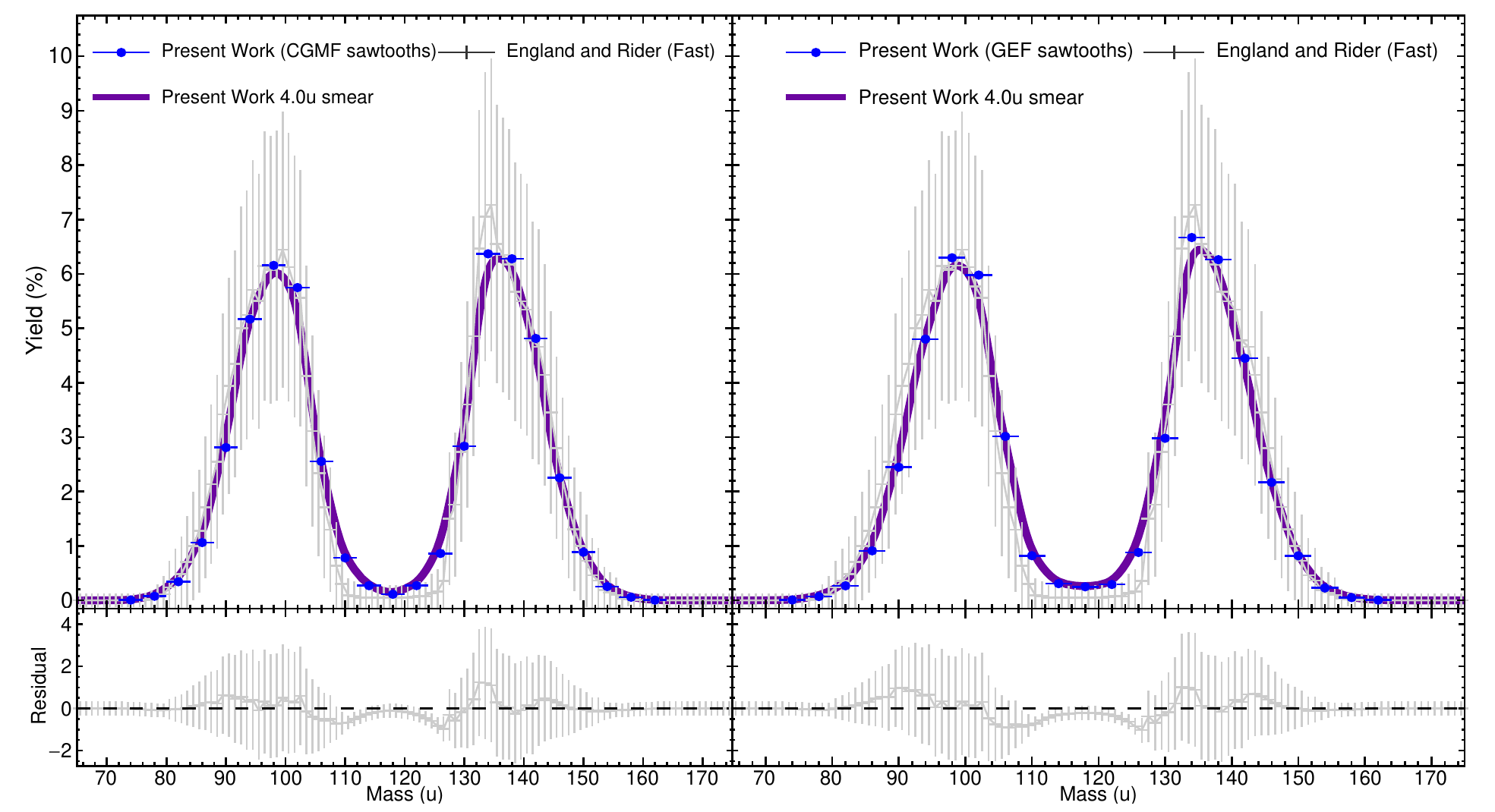}};
      \node[anchor=center, inner sep=0, label={[align=left, xshift=-0.5cm, yshift=2.575cm](a)}] at (yield.center) {};
      \node[anchor=center, inner sep=0, label={[align=left, xshift=6cm, yshift=2.575cm](b)}] at (yield.center) {};
      \node[anchor=center, inner sep=0, label={[align=left, xshift=-0.5cm, yshift=-2.375cm](c)}] at (yield.center) {};
      \node[anchor=center, inner sep=0, label={[align=left, xshift=6cm, yshift=-2.375cm](d)}] at (yield.center) {};
    \end{scope}
  \end{tikzpicture}
  \caption{\label{fig:ERfastYields} Top row - comparison of post-neutron evaporation mass yields for the present work (blue circles) obtained using CGMF neutron sawtooths (a) and GEF sawtooths (b) compared to the England and Rider evaluation\cite{England1993, Brown2018} for fast (0.5 MeV) incident neutrons (solid gray line and gray crosses) with $1-\sigma$ uncertainties. The solid purple line is the present data with a 4u FWHM smearing. (c) and (d) - residuals between present data with 4u smearing and the data points from the England and Rider evaluation.}
\end{figure*}
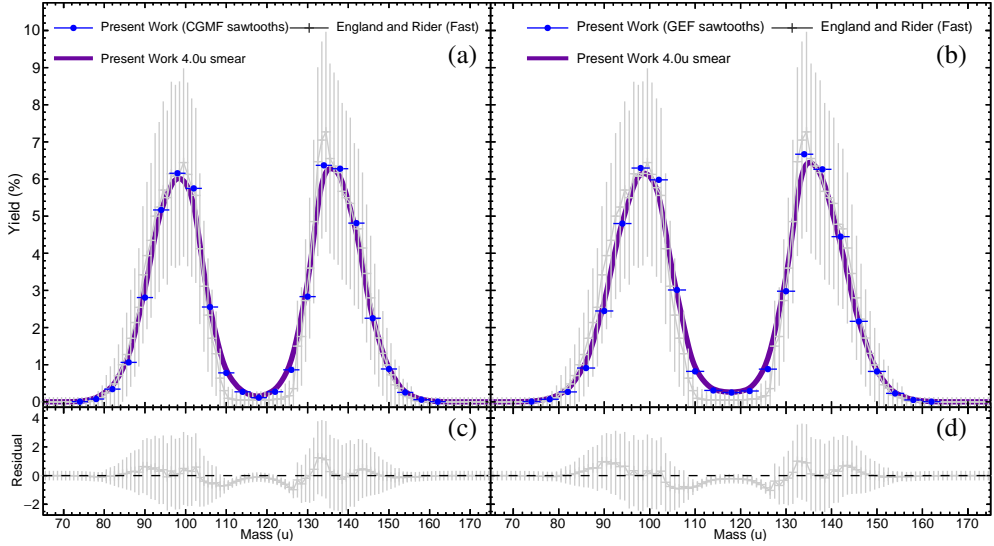
\subsection{Mass distributions}
\begin{figure*}[t]
  \begin{tikzpicture}[scale=1]
    \begin{scope}
      \node[anchor=south west, inner sep=0](yield) at (0,0) {\includegraphics[trim={0.35cm 0.05cm 0.375cm 0.2cm}, clip, width=\textwidth]{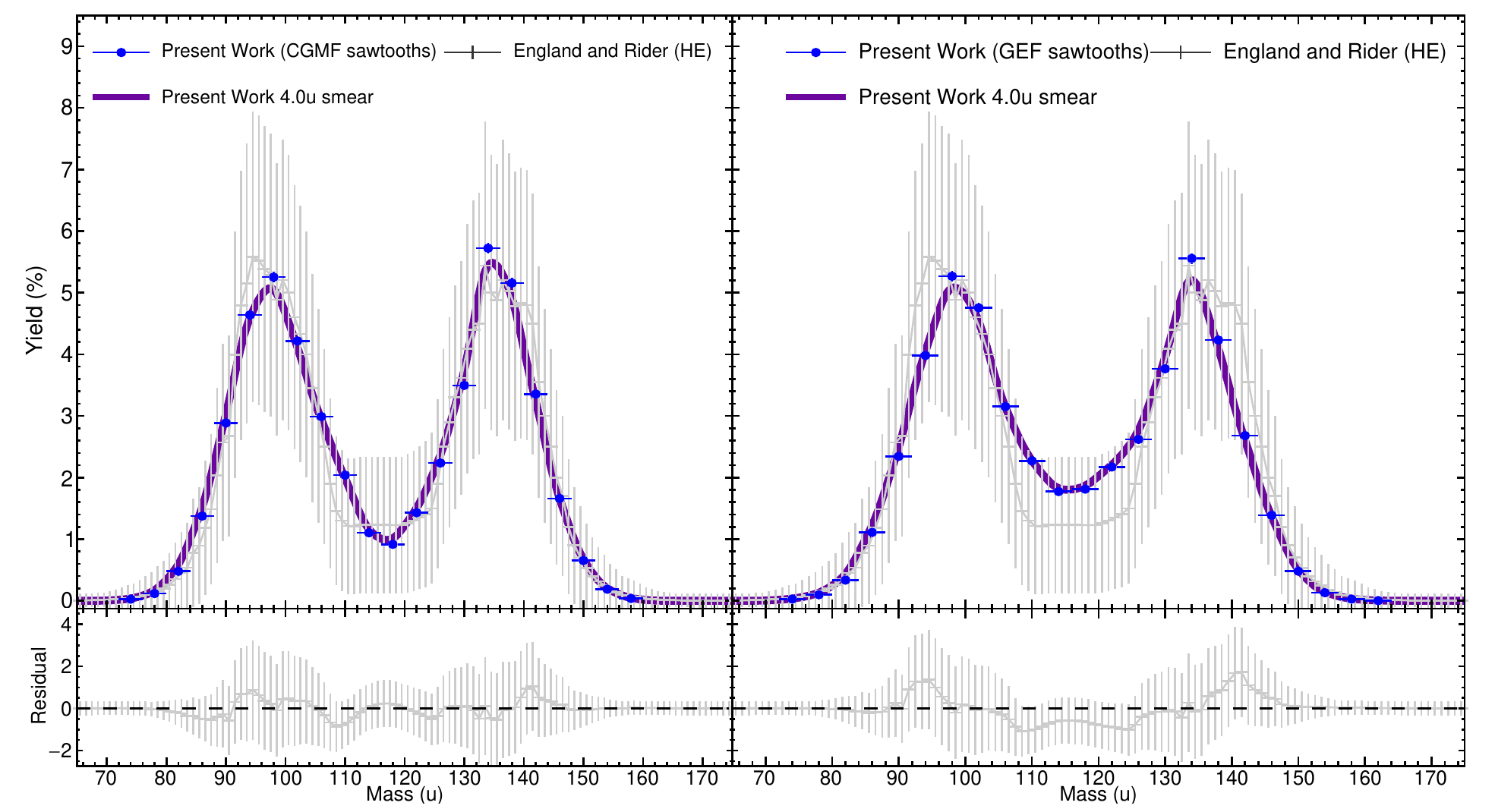}};
      \node[anchor=center, inner sep=0, label={[align=left, xshift=-0.5cm, yshift=2.575cm](a)}] at (yield.center) {};
      \node[anchor=center, inner sep=0, label={[align=left, xshift=6cm, yshift=2.575cm](b)}] at (yield.center) {};
      \node[anchor=center, inner sep=0, label={[align=left, xshift=-0.5cm, yshift=-2.375cm](c)}] at (yield.center) {};
      \node[anchor=center, inner sep=0, label={[align=left, xshift=6cm, yshift=-2.375cm](d)}] at (yield.center) {};
    \end{scope}
  \end{tikzpicture}
  \caption{\label{fig:ERHEYields}Top row - Comparison of post-neutron evaporation mass yields for the present work (blue circles) obtained using CGMF neutron sawtooths (a) and GEF sawtooths (b) compared to the England and Rider evaluation\cite{England1993, Brown2018} for high energy (14 MeV) incident neutrons (solid gray line and gray crosses) with $1-\sigma$ uncertainties. The solid purple line is the present data with a 4u FWHM smearing. (c) and (d) - residuals between the present data with 4u smearing and the data points from the England and Rider evaluation.}
\end{figure*}
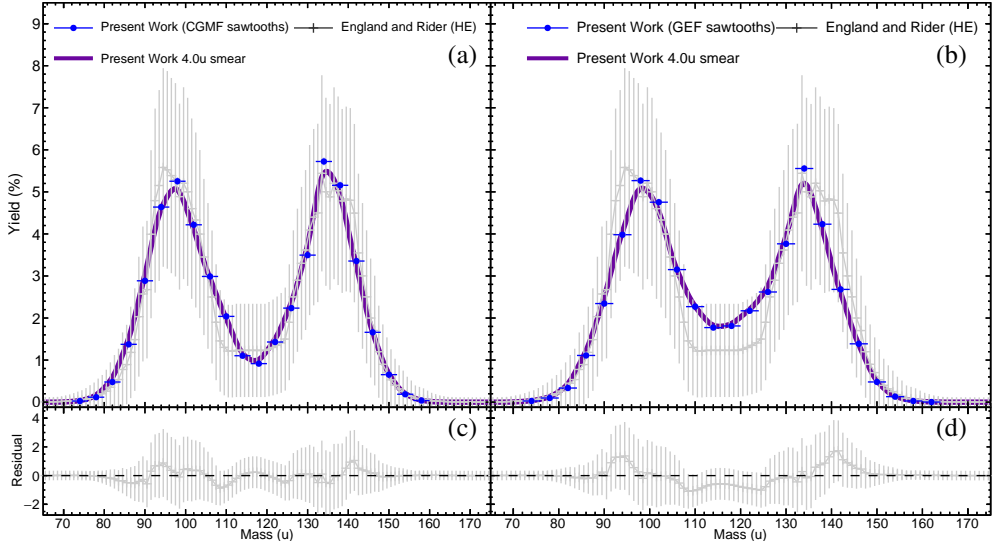

Pre- and Post-neutron emission mass yield distributions were extracted from the data via the $2E$ method using the neutron sawtooths from both the CGMF and GEF models. The post-neutron evaporation mass yields obtained from both models are compared to the England and Rider \cite{England1993,Brown2018} evaluation for ``fast'' neutrons (viz.\ - an incident neutron energy of $E_n = 0.5$~MeV) in \cref{fig:ERfastYields} and ``high-energy'' neutrons (viz.\ - an incident neutron energy of $E_n = 14$~MeV) in \cref{fig:ERHEYields}. The mass yields in both \cref{fig:ERfastYields,fig:ERHEYields} are normalized to 200\%. Events with incident neutron energies less than 1.0~MeV were selected to obtain the mass distributions at ``fast'' neutron energies (\cref{fig:ERfastYields}), whereas events with incident neutron energies in the range $E_n = 13.5 - 15.5$~MeV were selected to obtain the mass distributions for the ``high energy'' range (\cref{fig:ERHEYields}). The mass distributions extracted by utilizing the CGMF neutron sawtooths in the $2E$ analysis are presented in panels (a) and (b) of \cref{fig:ERfastYields,fig:ERHEYields}, whereas the mass distributions extracted by utilizing the GEF neutron sawtooths in the $2E$ analysis are presented in panels (c) and (d) of \cref{fig:ERfastYields,fig:ERHEYields}. The blue circles in \cref{fig:ERfastYields,fig:ERHEYields} are the mass yields extracted from the present experimental data, the black crosses connected by a solid black line are the raw England and Rider values with $1-\sigma$ uncertainty bars stated by England and Rider \cite{England1993,Brown2018}. The solid purple curves are the result of convolving a Gaussian with a FWHM of 4u with the measured mass yields. The bottom rows of \cref{fig:ERfastYields,fig:ERHEYields} show the residuals between extracted mass yield with 4u smearing and the data points from the England and Rider Evaluation.\\
\begin{table}[b]
\caption{\label{tab:chi2} $\chi^2$ values for the curves and data presented in \cref{fig:ERfastYields,fig:ERHEYields}. See text for details.}
\begin{indented}
\item[] 
\begin{tabular*}{0.33\textwidth}{@{\extracolsep{\fill}} llr}
   \br
   $E_n$    & Model   & $\chi^2$\\
   \midrule
    \multirow{2}{*}{fast}    & CGMF    & 0.38 \\
        & GEF     & 0.49 \\
    \multirow{2}{*}{HE}        & CGMF      & 0.05 \\
        & GEF     & 0.14 \\
   \br
 \end{tabular*}
\end{indented}
\end{table}
\indent The present results obtained using the CGMF neutron sawtooths for both incident neutron energies agrees with the England and Rider evaluation slightly better than those obtained using the neutron sawtooths from GEF. This is evident from the $\chi^2$ values between the present data with 4u smearing and the England and Rider data points, which are presented in \cref{tab:chi2}. The reader will note that the $1-\sigma$ uncertainties on the England and Rider evaluation for both the fast neutron energy mass yields and the high neutron energy mass yields are quite large, hence the relatively small $\chi^2$ values.
\section{Conclusion \label{sec:conclusion}}
Neutron induced fission of \np{237} was studied over a broad range of incident neutron energies from $E_n = 0.2 - 100$~MeV using a twin Frisch-gridded ionization chamber and a thin \np{237} target. Fission product energies and polar angles were measured for coincident fission products and pre- and post-neutron evaporation \tke\ and mass yields were deduced using the $2E$ method. Pre- and post-neutron emission mass yields were extracted. The present \tke\ results agree with previous measurements in the range $E_n = 0.2 - 5.5$~MeV. The present \tke\ data was also compared to a recent $2E$ measurement \cite{Pica2020} using an alternative measurement technique. The present data agrees with that measurement over the incident neutron energy range of $E_n = 0.2 - 100$~MeV with a significant reduction of the uncertainty in \tke\ as well as a much finer neutron energy binning. A flattening of the \tke\ relative to the GEF prediction is observed beginning at incident neutron energies of $E_n \approx 40$~MeV, corroborating previous observations \cite{King2017, Yanez2018, Chemey2020, Pica2020} in this energy range. However, the interpretation of this observation is ambiguous, as pre-equilibrium, pre-scission phenomena (such as fission induced by inelastic neutron scattering) as well as incomplete momentum transfer (both of which are neglected in the $2E$ analysis of this and previous studies), likely have a significant impact on the measurement at these energies.
\ack
The authors are grateful to Walter Loveland and his team at Oregon State University for the preparation of nuclear materials used in this experiment. This work was supported by and performed under the auspices of the US Department of Energy through the Los Alamos National Laboratory. Los Alamos National Laboratory is operated by Triad National Security, LLC, for the National Nuclear Security Administration of the U.S. Department of Energy (Contract No.~$89233218CNA000001$).\\
\appendix
\section*{Appendix}
\setcounter{section}{1}
\setlength{\extrarowheight}{0.25em}
{\fontsize{6}{7}\selectfont
  \begin{longtable}{@{\extracolsep{0.25em}}W{l}{0.06\textwidth} W{l}{0.07\textwidth} W{c}{0.06\textwidth} W{c}{0.07\textwidth} W{r}{0.06\textwidth} W{l}{0.07\textwidth} W{c}{0.06\textwidth} W{c}{0.07\textwidth} W{r}{0.06\textwidth} W{r}{0.08\textwidth}}
\caption{\label{tab:tke}Pre- and post-neutron evaporation $\langle TKE\rangle$ and $\sigma_{\mathrm{TKE}}$ values for given incident neutron energies obtained using both the GEF and CGMF neutron sawtooths in the $2E$ analysis. All units are in MeV.}\\[-6pt]
\br
  & \multicolumn{4}{c}{GEF} & \multicolumn{4}{c}{CGMF} & \\
  \cline{2-5} \cline{6-9} \\
  $E_n$  & $\langle TKE\rangle_{\mathrm{post}}$  & $\sigma_{\mathrm{post}}$  & $\langle TKE\rangle_{\mathrm{pre}}$  & $\sigma_{\mathrm{pre}}$  & $\langle TKE\rangle_{\mathrm{post}}$  & $\sigma_{\mathrm{post}}$  & $\langle TKE\rangle_{\mathrm{pre}}$  & $\sigma_{\mathrm{pre}}$  & $N_{\mathrm{events}}$ \\
\mr
$0.31_{- 0.09}^{+0.11}$ 	 & $171.03 \pm 0.12$ 	 & $11.94 \pm 0.08$ 	 & $172.93 \pm 0.12$ 	 & $12.05 \pm 0.09$ 	 & $171.02 \pm 0.12$ 	 & $11.93 \pm 0.08$ 	 & $173.18 \pm 0.12$ 	 & $12.06 \pm 0.09$ 	 & $10034 \pm 100.17$ \\
$0.48_{- 0.06}^{+0.05}$ 	 & $172.37 \pm 0.11$ 	 & $11.21 \pm 0.07$ 	 & $174.26 \pm 0.11$ 	 & $11.33 \pm 0.08$ 	 & $172.34 \pm 0.11$ 	 & $11.21 \pm 0.07$ 	 & $174.51 \pm 0.11$ 	 & $11.34 \pm 0.08$ 	 & $11194 \pm 105.80$ \\
$0.56\pm0.03$ 	             & $172.28 \pm 0.11$ 	 & $11.41 \pm 0.08$ 	 & $174.16 \pm 0.11$ 	 & $11.53 \pm 0.08$ 	 & $172.25 \pm 0.11$ 	 & $11.41 \pm 0.08$ 	 & $174.42 \pm 0.11$ 	 & $11.54 \pm 0.08$ 	 & $10268 \pm 101.33$ \\
$0.62_{- 0.03}^{+0.02}$ 	 & $172.27 \pm 0.11$ 	 & $11.45 \pm 0.08$ 	 & $174.16 \pm 0.11$ 	 & $11.58 \pm 0.08$ 	 & $172.25 \pm 0.11$ 	 & $11.45 \pm 0.08$ 	 & $174.43 \pm 0.11$ 	 & $11.58 \pm 0.08$ 	 & $11222 \pm 105.93$ \\
$0.66\pm0.02$ 	             & $172.47 \pm 0.11$ 	 & $11.23 \pm 0.08$ 	 & $174.37 \pm 0.11$ 	 & $11.35 \pm 0.08$ 	 & $172.45 \pm 0.11$ 	 & $11.23 \pm 0.08$ 	 & $174.64 \pm 0.11$ 	 & $11.35 \pm 0.08$ 	 & $10091 \pm 100.45$ \\
$0.70\pm0.02$ 	             & $172.32 \pm 0.11$ 	 & $11.38 \pm 0.08$ 	 & $174.21 \pm 0.11$ 	 & $11.50 \pm 0.08$ 	 & $172.29 \pm 0.11$ 	 & $11.38 \pm 0.08$ 	 & $174.48 \pm 0.11$ 	 & $11.50 \pm 0.08$ 	 & $11157 \pm 105.63$ \\
$0.74\pm0.02$ 	             & $172.38 \pm 0.11$ 	 & $11.35 \pm 0.07$ 	 & $174.28 \pm 0.11$ 	 & $11.47 \pm 0.08$ 	 & $172.36 \pm 0.11$ 	 & $11.35 \pm 0.07$ 	 & $174.55 \pm 0.11$ 	 & $11.47 \pm 0.08$ 	 & $11524 \pm 107.35$ \\
$0.78\pm0.02$ 	             & $172.25 \pm 0.11$ 	 & $11.48 \pm 0.07$ 	 & $174.15 \pm 0.11$ 	 & $11.60 \pm 0.08$ 	 & $172.23 \pm 0.11$ 	 & $11.47 \pm 0.07$ 	 & $174.43 \pm 0.11$ 	 & $11.60 \pm 0.08$ 	 & $11917 \pm 109.17$ \\
$0.82\pm0.02$ 	             & $172.38 \pm 0.10$ 	 & $11.15 \pm 0.07$ 	 & $174.28 \pm 0.10$ 	 & $11.27 \pm 0.07$ 	 & $172.36 \pm 0.10$ 	 & $11.14 \pm 0.07$ 	 & $174.57 \pm 0.10$ 	 & $11.26 \pm 0.07$ 	 & $12404 \pm 111.37$ \\
$0.86\pm0.02$ 	             & $172.49 \pm 0.10$ 	 & $11.27 \pm 0.07$ 	 & $174.40 \pm 0.10$ 	 & $11.40 \pm 0.07$ 	 & $172.46 \pm 0.10$ 	 & $11.27 \pm 0.07$ 	 & $174.67 \pm 0.10$ 	 & $11.40 \pm 0.07$ 	 & $12531 \pm 111.94$ \\
$0.90\pm0.02$ 	             & $172.49 \pm 0.10$ 	 & $11.29 \pm 0.07$ 	 & $174.39 \pm 0.10$ 	 & $11.41 \pm 0.07$ 	 & $172.46 \pm 0.10$ 	 & $11.28 \pm 0.07$ 	 & $174.68 \pm 0.10$ 	 & $11.41 \pm 0.07$ 	 & $12928 \pm 113.70$ \\
$0.94\pm0.02$ 	             & $172.52 \pm 0.10$ 	 & $11.30 \pm 0.07$ 	 & $174.43 \pm 0.10$ 	 & $11.43 \pm 0.07$ 	 & $172.50 \pm 0.10$ 	 & $11.30 \pm 0.07$ 	 & $174.72 \pm 0.10$ 	 & $11.42 \pm 0.07$ 	 & $13297 \pm 115.31$ \\
$0.98\pm0.02$ 	             & $172.56 \pm 0.10$ 	 & $11.15 \pm 0.07$ 	 & $174.48 \pm 0.10$ 	 & $11.27 \pm 0.07$ 	 & $172.53 \pm 0.10$ 	 & $11.15 \pm 0.07$ 	 & $174.76 \pm 0.10$ 	 & $11.27 \pm 0.07$ 	 & $12682 \pm 112.61$ \\
$1.02\pm0.02$ 	             & $172.56 \pm 0.10$ 	 & $11.22 \pm 0.07$ 	 & $174.48 \pm 0.10$ 	 & $11.34 \pm 0.07$ 	 & $172.54 \pm 0.10$ 	 & $11.22 \pm 0.07$ 	 & $174.76 \pm 0.10$ 	 & $11.35 \pm 0.07$ 	 & $12455 \pm 111.60$ \\
$1.08_{- 0.04}^{+0.03}$ 	 & $172.46 \pm 0.08$ 	 & $11.22 \pm 0.05$ 	 & $174.37 \pm 0.08$ 	 & $11.34 \pm 0.05$ 	 & $172.43 \pm 0.08$ 	 & $11.22 \pm 0.05$ 	 & $174.66 \pm 0.08$ 	 & $11.34 \pm 0.05$ 	 & $21578 \pm 146.89$ \\
$1.14_{- 0.03}^{+0.04}$ 	 & $172.36 \pm 0.08$ 	 & $11.27 \pm 0.06$ 	 & $174.27 \pm 0.08$ 	 & $11.39 \pm 0.06$ 	 & $172.33 \pm 0.08$ 	 & $11.26 \pm 0.06$ 	 & $174.56 \pm 0.08$ 	 & $11.39 \pm 0.06$ 	 & $20915 \pm 144.62$ \\
$1.21_{- 0.03}^{+0.04}$ 	 & $172.33 \pm 0.08$ 	 & $11.33 \pm 0.06$ 	 & $174.24 \pm 0.08$ 	 & $11.45 \pm 0.06$ 	 & $172.30 \pm 0.08$ 	 & $11.33 \pm 0.06$ 	 & $174.53 \pm 0.08$ 	 & $11.45 \pm 0.06$ 	 & $20420 \pm 142.90$ \\
$1.29\pm0.04$ 	             & $172.36 \pm 0.08$ 	 & $11.36 \pm 0.05$ 	 & $174.31 \pm 0.08$ 	 & $11.48 \pm 0.06$ 	 & $172.33 \pm 0.08$ 	 & $11.35 \pm 0.05$ 	 & $174.61 \pm 0.08$ 	 & $11.48 \pm 0.06$ 	 & $21769 \pm 147.54$ \\
$1.37\pm0.04$ 	             & $172.38 \pm 0.08$ 	 & $11.26 \pm 0.05$ 	 & $174.32 \pm 0.08$ 	 & $11.38 \pm 0.06$ 	 & $172.35 \pm 0.08$ 	 & $11.26 \pm 0.05$ 	 & $174.63 \pm 0.08$ 	 & $11.38 \pm 0.06$ 	 & $21167 \pm 145.49$ \\
$1.45\pm0.04$ 	             & $172.30 \pm 0.08$ 	 & $11.18 \pm 0.06$ 	 & $174.25 \pm 0.08$ 	 & $11.31 \pm 0.06$ 	 & $172.27 \pm 0.08$ 	 & $11.18 \pm 0.06$ 	 & $174.55 \pm 0.08$ 	 & $11.31 \pm 0.06$ 	 & $20220 \pm 142.20$ \\
$1.53_{- 0.04}^{+0.05}$ 	 & $172.19 \pm 0.08$ 	 & $11.38 \pm 0.05$ 	 & $174.14 \pm 0.08$ 	 & $11.50 \pm 0.06$ 	 & $172.16 \pm 0.08$ 	 & $11.37 \pm 0.05$ 	 & $174.44 \pm 0.08$ 	 & $11.50 \pm 0.06$ 	 & $21670 \pm 147.21$ \\
$1.62_{- 0.04}^{+0.05}$ 	 & $172.28 \pm 0.08$ 	 & $11.22 \pm 0.06$ 	 & $174.23 \pm 0.08$ 	 & $11.35 \pm 0.06$ 	 & $172.25 \pm 0.08$ 	 & $11.22 \pm 0.06$ 	 & $174.53 \pm 0.08$ 	 & $11.35 \pm 0.06$ 	 & $20812 \pm 144.26$ \\
$1.71_{- 0.04}^{+0.05}$ 	 & $172.21 \pm 0.08$ 	 & $11.28 \pm 0.06$ 	 & $174.16 \pm 0.08$ 	 & $11.40 \pm 0.06$ 	 & $172.19 \pm 0.08$ 	 & $11.27 \pm 0.06$ 	 & $174.47 \pm 0.08$ 	 & $11.40 \pm 0.06$ 	 & $20162 \pm 141.99$ \\
$1.81\pm0.05$ 	             & $172.14 \pm 0.08$ 	 & $11.25 \pm 0.05$ 	 & $174.13 \pm 0.08$ 	 & $11.37 \pm 0.06$ 	 & $172.10 \pm 0.08$ 	 & $11.25 \pm 0.05$ 	 & $174.44 \pm 0.08$ 	 & $11.38 \pm 0.06$ 	 & $21074 \pm 145.17$ \\
$1.91_{- 0.05}^{+0.06}$ 	 & $172.25 \pm 0.08$ 	 & $11.38 \pm 0.05$ 	 & $174.24 \pm 0.08$ 	 & $11.51 \pm 0.06$ 	 & $172.21 \pm 0.08$ 	 & $11.38 \pm 0.05$ 	 & $174.55 \pm 0.08$ 	 & $11.52 \pm 0.06$ 	 & $21529 \pm 146.73$ \\
$2.02_{- 0.05}^{+0.06}$ 	 & $172.17 \pm 0.08$ 	 & $11.22 \pm 0.06$ 	 & $174.16 \pm 0.08$ 	 & $11.34 \pm 0.06$ 	 & $172.14 \pm 0.08$ 	 & $11.21 \pm 0.06$ 	 & $174.47 \pm 0.08$ 	 & $11.35 \pm 0.06$ 	 & $20783 \pm 144.16$ \\
$2.14\pm0.06$ 	             & $171.97 \pm 0.08$ 	 & $11.33 \pm 0.05$ 	 & $173.96 \pm 0.08$ 	 & $11.46 \pm 0.06$ 	 & $171.94 \pm 0.08$ 	 & $11.33 \pm 0.05$ 	 & $174.27 \pm 0.08$ 	 & $11.46 \pm 0.06$ 	 & $21543 \pm 146.78$ \\
$2.26\pm0.06$ 	             & $172.15 \pm 0.08$ 	 & $11.22 \pm 0.06$ 	 & $174.16 \pm 0.08$ 	 & $11.35 \pm 0.06$ 	 & $172.11 \pm 0.08$ 	 & $11.22 \pm 0.06$ 	 & $174.48 \pm 0.08$ 	 & $11.35 \pm 0.06$ 	 & $20143 \pm 141.93$ \\
$2.38_{- 0.06}^{+0.07}$ 	 & $171.99 \pm 0.08$ 	 & $11.26 \pm 0.06$ 	 & $174.02 \pm 0.08$ 	 & $11.39 \pm 0.06$ 	 & $171.95 \pm 0.08$ 	 & $11.26 \pm 0.05$ 	 & $174.34 \pm 0.08$ 	 & $11.40 \pm 0.06$ 	 & $20980 \pm 144.84$ \\
$2.52\pm0.07$ 	             & $172.00 \pm 0.08$ 	 & $11.31 \pm 0.06$ 	 & $174.04 \pm 0.08$ 	 & $11.44 \pm 0.06$ 	 & $171.97 \pm 0.08$ 	 & $11.31 \pm 0.06$ 	 & $174.35 \pm 0.08$ 	 & $11.45 \pm 0.06$ 	 & $20641 \pm 143.67$ \\
$2.66_{- 0.07}^{+0.08}$ 	 & $171.96 \pm 0.08$ 	 & $11.25 \pm 0.06$ 	 & $174.00 \pm 0.08$ 	 & $11.38 \pm 0.06$ 	 & $171.93 \pm 0.08$ 	 & $11.25 \pm 0.06$ 	 & $174.32 \pm 0.08$ 	 & $11.39 \pm 0.06$ 	 & $20924 \pm 144.65$ \\
$2.82\pm0.08$ 	             & $171.89 \pm 0.08$ 	 & $11.25 \pm 0.06$ 	 & $173.96 \pm 0.08$ 	 & $11.38 \pm 0.06$ 	 & $171.85 \pm 0.08$ 	 & $11.25 \pm 0.06$ 	 & $174.29 \pm 0.08$ 	 & $11.39 \pm 0.06$ 	 & $20639 \pm 143.66$ \\
$2.98_{- 0.08}^{+0.09}$ 	 & $171.71 \pm 0.08$ 	 & $11.26 \pm 0.06$ 	 & $173.78 \pm 0.08$ 	 & $11.39 \pm 0.06$ 	 & $171.68 \pm 0.08$ 	 & $11.26 \pm 0.06$ 	 & $174.12 \pm 0.08$ 	 & $11.40 \pm 0.06$ 	 & $20624 \pm 143.61$ \\
$3.16_{- 0.09}^{+0.10}$ 	 & $171.64 \pm 0.08$ 	 & $11.34 \pm 0.06$ 	 & $173.71 \pm 0.08$ 	 & $11.47 \pm 0.06$ 	 & $171.61 \pm 0.08$ 	 & $11.34 \pm 0.06$ 	 & $174.05 \pm 0.08$ 	 & $11.48 \pm 0.06$ 	 & $20886 \pm 144.52$ \\
$3.36\pm0.10$ 	             & $171.60 \pm 0.08$ 	 & $11.36 \pm 0.06$ 	 & $173.70 \pm 0.08$ 	 & $11.49 \pm 0.06$ 	 & $171.56 \pm 0.08$ 	 & $11.35 \pm 0.06$ 	 & $174.06 \pm 0.08$ 	 & $11.50 \pm 0.06$ 	 & $20367 \pm 142.71$ \\
$3.57\pm0.11$ 	             & $171.41 \pm 0.08$ 	 & $11.26 \pm 0.06$ 	 & $173.52 \pm 0.08$ 	 & $11.40 \pm 0.06$ 	 & $171.38 \pm 0.08$ 	 & $11.26 \pm 0.06$ 	 & $173.88 \pm 0.08$ 	 & $11.40 \pm 0.06$ 	 & $20215 \pm 142.18$ \\
$3.80\pm0.12$ 	             & $171.38 \pm 0.08$ 	 & $11.17 \pm 0.06$ 	 & $173.52 \pm 0.08$ 	 & $11.30 \pm 0.06$ 	 & $171.34 \pm 0.08$ 	 & $11.17 \pm 0.06$ 	 & $173.88 \pm 0.08$ 	 & $11.31 \pm 0.06$ 	 & $20390 \pm 142.79$ \\
$4.05\pm0.13$ 	             & $171.27 \pm 0.08$ 	 & $11.28 \pm 0.06$ 	 & $173.42 \pm 0.08$ 	 & $11.42 \pm 0.06$ 	 & $171.23 \pm 0.08$ 	 & $11.28 \pm 0.06$ 	 & $173.78 \pm 0.08$ 	 & $11.43 \pm 0.06$ 	 & $20048 \pm 141.59$ \\
$4.32_{- 0.14}^{+0.15}$ 	 & $171.24 \pm 0.08$ 	 & $11.30 \pm 0.06$ 	 & $173.41 \pm 0.08$ 	 & $11.44 \pm 0.06$ 	 & $171.20 \pm 0.08$ 	 & $11.29 \pm 0.06$ 	 & $173.80 \pm 0.08$ 	 & $11.44 \pm 0.06$ 	 & $20149 \pm 141.95$ \\
$4.63\pm0.16$ 	             & $171.09 \pm 0.08$ 	 & $11.25 \pm 0.06$ 	 & $173.27 \pm 0.08$ 	 & $11.40 \pm 0.06$ 	 & $171.04 \pm 0.08$ 	 & $11.25 \pm 0.06$ 	 & $173.66 \pm 0.08$ 	 & $11.40 \pm 0.06$ 	 & $20203 \pm 142.14$ \\
$4.97_{- 0.18}^{+0.19}$ 	 & $170.93 \pm 0.08$ 	 & $11.31 \pm 0.06$ 	 & $173.15 \pm 0.08$ 	 & $11.46 \pm 0.06$ 	 & $170.88 \pm 0.08$ 	 & $11.31 \pm 0.06$ 	 & $173.55 \pm 0.08$ 	 & $11.47 \pm 0.06$ 	 & $20429 \pm 142.93$ \\
$5.36_{- 0.20}^{+0.22}$ 	 & $170.65 \pm 0.08$ 	 & $11.29 \pm 0.06$ 	 & $172.88 \pm 0.08$ 	 & $11.45 \pm 0.06$ 	 & $170.60 \pm 0.08$ 	 & $11.29 \pm 0.06$ 	 & $173.30 \pm 0.08$ 	 & $11.45 \pm 0.06$ 	 & $20104 \pm 141.79$ \\
$5.81\pm0.23$ 	             & $170.58 \pm 0.08$ 	 & $11.38 \pm 0.06$ 	 & $172.80 \pm 0.08$ 	 & $11.53 \pm 0.06$ 	 & $170.54 \pm 0.08$ 	 & $11.38 \pm 0.06$ 	 & $173.29 \pm 0.08$ 	 & $11.54 \pm 0.06$ 	 & $20013 \pm 141.47$ \\
$6.27_{- 0.23}^{+0.22}$ 	 & $170.78 \pm 0.08$ 	 & $11.43 \pm 0.06$ 	 & $172.96 \pm 0.08$ 	 & $11.58 \pm 0.06$ 	 & $170.74 \pm 0.08$ 	 & $11.43 \pm 0.06$ 	 & $173.46 \pm 0.08$ 	 & $11.59 \pm 0.06$ 	 & $20284 \pm 142.42$ \\
$6.72_{- 0.23}^{+0.22}$ 	 & $170.75 \pm 0.08$ 	 & $11.37 \pm 0.06$ 	 & $172.92 \pm 0.08$ 	 & $11.51 \pm 0.06$ 	 & $170.72 \pm 0.08$ 	 & $11.36 \pm 0.06$ 	 & $173.37 \pm 0.08$ 	 & $11.51 \pm 0.06$ 	 & $20151 \pm 141.95$ \\
$7.17_{- 0.23}^{+0.24}$ 	 & $170.79 \pm 0.08$ 	 & $11.45 \pm 0.06$ 	 & $172.97 \pm 0.08$ 	 & $11.59 \pm 0.06$ 	 & $170.75 \pm 0.08$ 	 & $11.45 \pm 0.06$ 	 & $173.40 \pm 0.08$ 	 & $11.59 \pm 0.06$ 	 & $20052 \pm 141.61$ \\
$7.67_{- 0.26}^{+0.27}$ 	 & $170.63 \pm 0.08$ 	 & $11.46 \pm 0.06$ 	 & $172.84 \pm 0.08$ 	 & $11.61 \pm 0.06$ 	 & $170.59 \pm 0.08$ 	 & $11.46 \pm 0.06$ 	 & $173.26 \pm 0.08$ 	 & $11.60 \pm 0.06$ 	 & $20276 \pm 142.39$ \\
$8.23_{- 0.29}^{+0.31}$ 	 & $170.46 \pm 0.08$ 	 & $11.50 \pm 0.06$ 	 & $172.72 \pm 0.08$ 	 & $11.65 \pm 0.06$ 	 & $170.42 \pm 0.08$ 	 & $11.49 \pm 0.06$ 	 & $173.15 \pm 0.08$ 	 & $11.65 \pm 0.06$ 	 & $20162 \pm 141.99$ \\
$8.89_{- 0.35}^{+0.37}$ 	 & $170.19 \pm 0.08$ 	 & $11.63 \pm 0.06$ 	 & $172.52 \pm 0.08$ 	 & $11.79 \pm 0.06$ 	 & $170.15 \pm 0.08$ 	 & $11.62 \pm 0.06$ 	 & $172.97 \pm 0.08$ 	 & $11.78 \pm 0.06$ 	 & $20101 \pm 141.78$ \\
$9.68_{- 0.42}^{+0.46}$ 	 & $169.81 \pm 0.08$ 	 & $11.67 \pm 0.06$ 	 & $172.19 \pm 0.08$ 	 & $11.83 \pm 0.06$ 	 & $169.76 \pm 0.08$ 	 & $11.66 \pm 0.06$ 	 & $172.70 \pm 0.08$ 	 & $11.83 \pm 0.06$ 	 & $20072 \pm 141.68$ \\
$10.40_{- 0.26}^{+0.28}$ 	 & $169.66 \pm 0.12$ 	 & $11.70 \pm 0.08$ 	 & $172.09 \pm 0.12$ 	 & $11.87 \pm 0.08$ 	 & $169.62 \pm 0.12$ 	 & $11.70 \pm 0.08$ 	 & $172.63 \pm 0.12$ 	 & $11.87 \pm 0.08$ 	 & $10171 \pm 100.85$ \\
$10.98\pm0.30$ 	             & $169.21 \pm 0.12$ 	 & $11.66 \pm 0.08$ 	 & $171.69 \pm 0.12$ 	 & $11.82 \pm 0.08$ 	 & $169.17 \pm 0.12$ 	 & $11.65 \pm 0.08$ 	 & $172.26 \pm 0.12$ 	 & $11.83 \pm 0.08$ 	 & $10033 \pm 100.16$ \\
$11.62_{- 0.34}^{+0.36}$ 	 & $168.99 \pm 0.12$ 	 & $11.82 \pm 0.08$ 	 & $171.51 \pm 0.12$ 	 & $12.00 \pm 0.08$ 	 & $168.96 \pm 0.12$ 	 & $11.81 \pm 0.08$ 	 & $172.11 \pm 0.12$ 	 & $12.00 \pm 0.08$ 	 & $10100 \pm 100.50$ \\
$12.36_{- 0.38}^{+0.41}$ 	 & $168.91 \pm 0.12$ 	 & $11.79 \pm 0.08$ 	 & $171.46 \pm 0.12$ 	 & $11.96 \pm 0.08$ 	 & $168.87 \pm 0.12$ 	 & $11.78 \pm 0.08$ 	 & $172.05 \pm 0.12$ 	 & $11.97 \pm 0.08$ 	 & $10049 \pm 100.24$ \\
$13.22_{- 0.45}^{+0.47}$ 	 & $168.47 \pm 0.12$ 	 & $11.74 \pm 0.08$ 	 & $171.04 \pm 0.12$ 	 & $11.92 \pm 0.08$ 	 & $168.43 \pm 0.12$ 	 & $11.74 \pm 0.08$ 	 & $171.56 \pm 0.12$ 	 & $11.92 \pm 0.08$ 	 & $10010 \pm 100.05$ \\
$14.19_{- 0.50}^{+0.52}$ 	 & $168.13 \pm 0.12$ 	 & $11.78 \pm 0.08$ 	 & $170.70 \pm 0.12$ 	 & $11.96 \pm 0.08$ 	 & $168.09 \pm 0.12$ 	 & $11.77 \pm 0.08$ 	 & $171.08 \pm 0.12$ 	 & $11.93 \pm 0.08$ 	 & $10029 \pm 100.14$ \\
$15.26_{- 0.55}^{+0.57}$ 	 & $168.21 \pm 0.12$ 	 & $11.61 \pm 0.08$ 	 & $171.04 \pm 0.12$ 	 & $11.81 \pm 0.08$ 	 & $168.22 \pm 0.12$ 	 & $11.61 \pm 0.08$ 	 & $171.16 \pm 0.12$ 	 & $11.76 \pm 0.08$ 	 & $10011 \pm 100.05$ \\
$16.43_{- 0.60}^{+0.63}$ 	 & $168.24 \pm 0.12$ 	 & $11.84 \pm 0.08$ 	 & $170.87 \pm 0.12$ 	 & $12.01 \pm 0.08$ 	 & $168.22 \pm 0.12$ 	 & $11.83 \pm 0.08$ 	 & $171.15 \pm 0.12$ 	 & $11.98 \pm 0.08$ 	 & $10049 \pm 100.24$ \\
$17.75_{- 0.69}^{+0.72}$ 	 & $168.01 \pm 0.12$ 	 & $11.91 \pm 0.08$ 	 & $170.69 \pm 0.12$ 	 & $12.09 \pm 0.09$ 	 & $167.99 \pm 0.12$ 	 & $11.91 \pm 0.08$ 	 & $170.96 \pm 0.12$ 	 & $12.06 \pm 0.09$ 	 & $10040 \pm 100.20$ \\
$19.28_{- 0.81}^{+0.83}$ 	 & $167.45 \pm 0.12$ 	 & $11.98 \pm 0.08$ 	 & $170.18 \pm 0.12$ 	 & $12.15 \pm 0.09$ 	 & $167.43 \pm 0.12$ 	 & $11.96 \pm 0.08$ 	 & $170.51 \pm 0.12$ 	 & $12.14 \pm 0.09$ 	 & $10052 \pm 100.26$ \\
$21.00_{- 0.89}^{+0.92}$	 & $167.16 \pm 0.12$	 & $11.89 \pm 0.08$	     & $169.91 \pm 0.12$	 & $12.08 \pm 0.09$	     & - & - & - & - & $10038 \pm 100.19$ \\
$22.89_{- 0.97}^{+1.02}$ 	 & $167.20 \pm 0.12$ 	 & $11.97 \pm 0.08$ 	 & $170.01 \pm 0.12$ 	 & $12.16 \pm 0.09$ 	 & - & - & - & - & $10038 \pm 100.19$ \\
$25.00_{- 1.09}^{+1.12}$ 	 & $166.99 \pm 0.12$ 	 & $12.04 \pm 0.09$ 	 & $169.84 \pm 0.12$ 	 & $12.23 \pm 0.09$ 	 & - & - & - & - & $10035 \pm 100.17$ \\
$27.33_{- 1.21}^{+1.27}$ 	 & $166.62 \pm 0.12$ 	 & $11.98 \pm 0.08$ 	 & $169.52 \pm 0.12$ 	 & $12.17 \pm 0.09$ 	 & - & - & - & - & $10023 \pm 100.11$ \\
$29.95_{- 1.35}^{+1.42}$ 	 & $166.18 \pm 0.12$ 	 & $11.90 \pm 0.08$ 	 & $169.16 \pm 0.12$ 	 & $12.10 \pm 0.09$ 	 & - & - & - & - & $10029 \pm 100.14$ \\
$32.89_{- 1.52}^{+1.57}$ 	 & $166.19 \pm 0.12$ 	 & $11.95 \pm 0.08$ 	 & $169.20 \pm 0.12$ 	 & $12.18 \pm 0.09$ 	 & - & - & - & - & $10017 \pm 100.08$ \\
$36.15_{- 1.69}^{+1.80}$ 	 & $165.99 \pm 0.12$ 	 & $11.99 \pm 0.08$ 	 & $169.05 \pm 0.12$ 	 & $12.21 \pm 0.09$ 	 & - & - & - & - & $10012 \pm 100.06$ \\
$39.92_{- 1.97}^{+2.06}$ 	 & $165.74 \pm 0.12$ 	 & $12.10 \pm 0.09$ 	 & $168.84 \pm 0.12$ 	 & $12.32 \pm 0.09$ 	 & - & - & - & - & $10007 \pm 100.03$ \\
$44.23_{- 2.25}^{+2.33}$ 	 & $165.82 \pm 0.12$ 	 & $11.91 \pm 0.08$ 	 & $168.96 \pm 0.12$ 	 & $12.14 \pm 0.09$ 	 & - & - & - & - & $10015 \pm 100.07$ \\
$49.14_{- 2.58}^{+2.67}$ 	 & $165.28 \pm 0.12$ 	 & $12.14 \pm 0.09$ 	 & $168.47 \pm 0.12$ 	 & $12.38 \pm 0.09$ 	 & - & - & - & - & $10014 \pm 100.07$ \\
$54.85_{- 3.04}^{+3.21}$ 	 & $165.33 \pm 0.12$ 	 & $12.01 \pm 0.08$ 	 & $168.57 \pm 0.12$ 	 & $12.25 \pm 0.09$ 	 & - & - & - & - & $10007 \pm 100.03$ \\
$61.68_{- 3.62}^{+3.92}$ 	 & $165.13 \pm 0.12$ 	 & $12.03 \pm 0.09$ 	 & $168.43 \pm 0.12$ 	 & $12.28 \pm 0.09$ 	 & - & - & - & - & $10006 \pm 100.03$ \\
$70.08_{- 4.48}^{+4.81}$ 	 & $164.86 \pm 0.12$ 	 & $12.16 \pm 0.09$ 	 & $168.20 \pm 0.12$ 	 & $12.43 \pm 0.09$ 	 & - & - & - & - & $10000 \pm 100.00$ \\
$80.57_{- 5.68}^{+6.33}$ 	 & $164.95 \pm 0.12$ 	 & $12.19 \pm 0.09$ 	 & $168.36 \pm 0.12$ 	 & $12.46 \pm 0.09$ 	 & - & - & - & - & $10001 \pm 100.00$ \\
$93.12_{- 6.22}^{+6.88}$ 	 & $164.51 \pm 0.14$ 	 & $12.40 \pm 0.10$ 	 & $167.99 \pm 0.14$ 	 & $12.69 \pm 0.10$ 	 & - & - & - & - & $8131 \pm 90.17$ \\
\br
\end{longtable}
}
\bibliography{Connolly_Np-237.bib}
\end{document}